\documentclass[prc,twocolumn,showpacs,preprintnumbers,amsmath,amssymb]{revtex4}
\usepackage{txfonts}
\usepackage{amssymb}
\usepackage{graphicx}

\usepackage{tipa}
\usepackage{dcolumn}
\usepackage{bm}
\usepackage{multirow}
\usepackage{booktabs}

\begin{document}
\title{Beta-decay spectroscopy of $^{27}$S}
\author{L. J. Sun$^{1,2,3}$}
\author{X. X. Xu$^{1,2}$}
 \email{xinxing@hku.hk}
\author{S. Q. Hou$^{4}$}
\author{C. J. Lin$^{2,5}$}
 \email{cjlin@ciae.ac.cn}
\author{J. Jos\'{e} $^{6,7}$}
 \email{jordi.jose@upc.edu}
\author{J. Lee$^{1}$}
 \email{jleehc@hku.hk}
\author{J. J. He$^{8,9}$}
\author{Z. H. Li$^{10}$}
\author{J. S. Wang$^{4}$}
\author{\\C. X. Yuan$^{11}$}
\author{D. X. Wang$^{2}$}
\author{H. Y. Wu$^{10}$}
\author{P. F. Liang$^{1}$}
\author{Y. Y. Yang$^{4}$}
\author{Y. H. Lam$^{4}$}
\author{P. Ma$^{4}$}
\author{F. F. Duan$^{12,4}$}
\author{Z. H. Gao$^{4,12}$}
\author{\\Q. Hu$^{4}$}
\author{Z. Bai$^{4}$}
\author{J. B. Ma$^{4}$}
\author{J. G. Wang$^{4}$}
\author{F. P. Zhong$^{5,2}$}
\author{C. G. Wu$^{10}$}
\author{D. W. Luo$^{10}$}
\author{Y. Jiang$^{10}$}
\author{Y. Liu$^{10}$}
\author{\\D. S. Hou$^{4}$}
\author{R. Li$^{4}$}
\author{N. R. Ma$^{2}$}
\author{W. H. Ma$^{4}$}
\author{G. Z. Shi$^{4}$}
\author{G. M. Yu$^{4}$}
\author{D. Patel $^{4}$}
\author{S. Y. Jin$^{4}$}
\author{Y. F. Wang$^{13,4}$}
\author{\\Y. C. Yu$^{13,4}$}
\author{Q. W. Zhou$^{14,4}$}
\author{P. Wang$^{14,4}$}
\author{L. Y. Hu$^{15}$}
\author{X. Wang$^{10}$}
\author{H. L. Zang$^{10}$}
\author{P. J. Li$^{1}$}
\author{Q. Q. Zhao$^{1}$}
\author{L. Yang$^{2}$}
\author{\\P. W. Wen$^{2}$}
\author{F. Yang$^{2}$}
\author{H. M. Jia$^{2}$}
\author{G. L. Zhang$^{16}$}
\author{M. Pan$^{16,2}$}
\author{X. Y. Wang$^{16}$}
\author{H. H. Sun$^{2}$}
\author{Z. G. Hu$^{4}$}
\author{\\R. F. Chen$^{4}$}
\author{M. L. Liu$^{4}$}
\author{W. Q. Yang$^{4}$}
\author{Y. M. Zhao$^{3}$}
\author{H. Q. Zhang$^{2}$}
\collaboration{RIBLL Collaboration}

\affiliation{\footnotesize$^1$Department of Physics, The University of Hong Kong, Hong Kong, China\\
$^2$Department of Nuclear Physics, China Institute of Atomic Energy, Beijing 102413, China\\
$^3$School of Physics and Astronomy, Shanghai Jiao Tong University, Shanghai 200240, China\\
$^4$Institute of Modern Physics, Chinese Academy of Sciences, Lanzhou 730000, China\\
$^5$College of Physics and Technology, Guangxi Normal University, Guilin 541004, China\\
$^6$Departament de F\'{i}sica, EEBE, Universitat Polit\'{e}cnica de Catalunya, Av./ Eduard Maristany 10, E-08930 Barcelona, Spain\\
$^7$Institut d¡¯Estudis Espacials de Catalunya (IEEC), Ed. Nexus-201, C/ Gran Capit\'{a} 2-4, E-08034 Barcelona, Spain\\
$^8$Key Laboratory of Optical Astronomy, National Astronomical Observatories, Chinese Academy of Sciences, Beijing 100012, China\\
$^9$University of Chinese Academy of Sciences, 100049 Beijing, China\\
$^{10}$State Key Laboratory of Nuclear Physics and Technology, School of Physics, Peking University, Beijing 100871, China\\
$^{11}$Sino-French Institute of Nuclear Engineering and Technology, Sun Yat-Sen University, Zhuhai 519082, China\\
$^{12}$School of Nuclear Science and Technology, Lanzhou University, Lanzhou 730000, China\\
$^{13}$School of Physics and Astronomy, Yunnan University, Kunming 650091, China\\
$^{14}$School of Physical Science and Technology, Southwest University, Chongqing 400044, China\\
$^{15}$Fundamental Science on Nuclear Safety and Simulation Technology Laboratory, Harbin Engineering University, Harbin 150001, China\\
$^{16}$School of Physics and Nuclear Energy Engineering, Beihang University, Beijing 100191, China\\
}

\date{\today}

\begin{abstract}
\begin{description}
\item[Background] Beta-decay spectroscopy provides valuable nuclear physics input for thermonuclear reaction rates of astrophysical interest and stringent test for shell-model theories far from the stability line.
\item[Purpose] The available decay properties of proton drip-line nucleus $^{27}$S are insufficient to constrain the properties of the key resonance in $^{26}$Si$(p,\gamma)^{27}$P reaction rate and probe the possible mirror asymmetry. The decay scheme of $^{27}$S is complicated and far from being understood, which has motivated but also presented substantial challenges for our experiment.
\item[Method] The $^{27}$S ions were implanted into a double-sided silicon strip detector array surrounded by the high-purity germanium detectors, where the $\beta$-delayed protons and $\gamma$ rays were measured simultaneously.
\item[Results] The precise half-life of $^{27}$S, the excitation energies, $\beta$-feeding intensities, log~$ft$ values, and $B$(GT) values for the states of $^{27}$P populated in the $\beta$ decay of $^{27}$S are determined. The improved spectroscopic properties including are compared to the mirror $\beta$ decay of $^{27}$Na and to the shell-model calculations using the recently-developed USD$^*$ interaction. The present work has expanded greatly on the previously established decay scheme of $^{27}$S.
\item[Conclusions] The precise mass excess of $^{27}$P, the energy and the ratio between $\gamma$ and proton partial widths of the $3/2^+$ resonance were obtained, thereby determining the $^{26}$Si$(p,\gamma)^{27}$P reaction rate based mainly on experimental constraints. The first experimental evidence for the observation of mirror asymmetries for the transitions in the decays of $^{27}$S and $^{27}$Na is also provided. The shell-model calculations with the Hamiltonians including the modifications on single-particle energies and two-body matrix elements related to the proton $1s_{1/2}$ orbit give a better description of the spectroscopic properties.
\end{description}
\end{abstract}

\maketitle

\section{Introduction}
The investigation of exotic nuclei lying far from the stability line has been one of the attractive topics of nuclear physics during the past few decades~\cite{Blank_PPNP2008,Borge_PS2013,Pfutzner_RMP2012}. It is preferable to extend the test of isospin symmetry to the limit of the drip line with the advent of more powerful radioactive-beam facilities. The isospin symmetry in the nuclear medium can be violated by Coulomb interaction and charge-dependent parts of nucleon-nucleon interaction. The difference in the $ft$ values for mirror $\beta$ transitions is one of the observable signatures of isospin-symmetry breaking, which is also referred to as mirror asymmetry. This phenomenon has been reported for transitions in the mirror $\beta$ decays of several $sd$-shell nuclei, such as 
$^{17}$Ne$\rightarrow^{17}$F~\cite{Borge_PLB1993,Ozawa_JPG1998} and $^{17}$N$\rightarrow^{17}$O~\cite{Tilley_NPA1993}, $^{20}$Mg$\rightarrow^{20}$Na~\cite{Lund_EPJA2016,Sun_PRC2017} and $^{20}$O$\rightarrow^{20}$F~\cite{Alburger_PRC1987}, $^{22}$Si$\rightarrow^{22}$Al and $^{22}$O$\rightarrow^{22}$F~\cite{Weissman_JPG2005}, $^{24}$Si$\rightarrow^{24}$Al~\cite{Ichikawa_JPConf2011} and $^{24}$Ne$\rightarrow^{24}$Na~\cite{McDonald_PR1969}, $^{26}$P$\rightarrow^{26}$Si~\cite{Perez-Loureiro_PRC2016} and $^{26}$Na$\rightarrow^{26}$Mg~\cite{Grinyer_PRC2005}. The degree of asymmetry preserves important information on the nuclear structure of the states involved. $\beta$-decay spectroscopic study has proved to be a powerful tool to obtain the structure information adjacent to the drip-line, which provides a reliable isospin-symmetry-breaking correction to superallowed Fermi $\beta$ decay~\cite{Towner_PRC2008,Towner_PRC2010,Hardy_PRC2015}, as well as an excellent and stringent test the accuracy of shell-model predictions far from the valley of stability. It is desirable to investigate the possible mirror asymmetry for the case of $^{27}$S$\rightarrow^{27}$P and $^{27}$Na$\rightarrow^{27}$Mg~\cite{Janiak_PRC2017}. To provide a better understanding of the nature of isospin-symmetry breaking, high precision measurements of $\beta$ decays should be extended to more nuclides. On the other hand, the accurate theoretical description of the possible origins of isospin-symmetry breaking within a microscopic model is complicated and is also a goal of the current efforts~\cite{Lam_PRC2013,Smirnova_PRC2017,Kaneko_PLB2017}.

In addition, we also have an astrophysical motivation for the $\beta$-decay study of $^{27}$S. The production and destruction of $^{26}$Si via proton radiative captures: $^{25}$Al($p,\gamma)^{26}$Si$(p,\gamma)^{27}$P, have impacts on the amount of the isomeric and ground state of $^{26}$Al~\cite{Jose_APJ1999,Prantzos_PR1996}, in which the latter is of outstanding importance in $\gamma$-ray astronomy and cosmochemistry~\cite{Diehl_N2006,Wang_AA2009}. Most of the radiative capture reactions occurring in novae and type I X-ray bursts (XRB) proceed by resonant capture through narrow, isolated, resonances in the product nuclei~\cite{Parikh_AIPA2014,Jose_NPA2006,Jose_JPG2007,Schatz_PPNP2011,Parikh_PPNP2013}. Nuclear reaction measurements have been the preferred method to obtain the astrophysical reaction rate~\cite{Caggiano_PRC2001,Moon_NPA2005,Jung_PRC2012,Togano_NPA2005,Togano_EPJA2006,Guo_PRC2006,Togano_PRC2011,Marganiec_PRC2016}. Nevertheless, nuclear decay measurement provides new insights on the resonances of astrophysical interest in comparison with the reaction experiments~\cite{Wrede_PP2015,Fynbo_JPG2017}. For the $^{25}$Al$(p,\gamma)^{26}$Si reaction, the properties of the $3^+$ key resonance in $^{25}$Al$(p,\gamma)^{26}$Si reaction were measured via the $\beta$-decay spectroscopy of $^{26}$P. It was found that up to 30\% of the galactic $^{26}$Al were contributed from classical novae with the new $^{25}$Al$(p,\gamma)^{26}$Si reaction rate adopted~\cite{Bennett_PRL2013,Schwartz_PRC2015,Perez-Loureiro_PRC2016}. Similarly, accurate $\beta$-decay spectroscopy of $^{27}$S can also be utilized as an alternative way to obtain valuable information such as the energies, partial widths, spins, and parities of the key proton resonances in $^{27}$P and thus set experimental constraints on the model prediction of the role of $^{26}$Si$(p,\gamma)^{27}$P reaction in nova and XRB nucleosynthesis.

The lightest sulfur isotope within the drip line, $^{27}$S, is five neutrons away from the last stable isotope of sulfur on the chart of nuclides, and is also predicted to be a proton-halo candidate~\cite{Brown_PLB1996,Ren_PRC1996,Chen_JPG1998,Gupta_JPG2002}. Borrel \textit{et al}. performed the first decay spectroscopy of $^{27}$S by implanting the ions into silicon detectors. $\beta$-delayed two-proton emission was observed, but no other decay channels were reported due to the low statistics and high contamination~\cite{Borrel_NPA1991}. Canchel \textit{et al}. measured the $^{27}$S decay by using a similar method, and high-energy $\beta$-delayed proton branches were found~\cite{Canchel_EPJA2001}. Recently, Janiak \textit{et al}. identified two low-energy proton transitions by using an optical time projection chamber~\cite{Janiak_PRC2017}. The large $\beta$-decay energy of $^{27}$S implies that many decay channels are open. The available information of the decay properties of is still quite insufficient as yet, and therefore motivates this experiments to search for new $\beta$-delayed particles and $\gamma$ rays. Evidently, all previous $\beta$-decay measurements of $^{27}$S were focused on the proton branch, whereas the $\gamma$-ray branch has rarely been addressed. The complicated decay scheme can be well reconstructed by measuring as many $\gamma$ rays and particles emitted in the decay as possible~\cite{Saastamoinen_PRC2011,Koldste_PRC2013}. As Janiak \textit{et al}. pointed out, their experiment of $^{27}$S based on a time projection chamber technique was preferred for the clear identification of decay channels and for the precise determination of their absolute branching ratios. The exact energies of the relevant resonances and the probability ratios of $\gamma$-to-proton emission can be established accurately using a complementary detector array, which refers to several silicon detectors surrounded by germanium detectors~\cite{Janiak_PRC2017}.

Silicon detector is essential for charged-particle detection and various silicon-detector arrays have been built and successfully commissioned to measure the multi-particle and multi-step decay modes expected in the nuclei near the proton-drip line~\cite{Matea_NIMA2009,McCleskey_NIMA2013,Lund_EPJA2016,Sun_NIMA2015,Wang_NST2018}. Several innovative new techniques and solutions such as printed circuit boards, cryogenic system, leading edge discrimination, front-back coincidence of DSSD, and energy calibration from an internal source were conceived and implemented on the bases of our previous decay measurements with an implantation method~\cite{Sun_CPL2015,Sun_NIMA2015,Sun_PRC2017,Xu_PLB2017,Wang_IJMPE2018,Wang_NST2018,Wang_EPJA2018} and complete-kinematics measurements~\cite{Lin_PRC2009,Xu_28P_PRC2010,Xu_18Ne_PRC2010,Xu_PLB2013}. In the present experiment, in order to reliably extract information about the very rare decay events from disturbances, a high signal-to-noise ratio, a large solid angle coverage, a broad dynamic range, and operation stability of the detection system were achieved by combining all the techniques. The emitted particles and $\gamma$ rays in the $\beta$ decay of $^{27}$S were measured simultaneously with high efficiency and high energy resolution. A comprehensive decay scheme of $^{27}$S is constructed and compared to theoretical calculations and to the decay of the mirror nucleus.

\section{Experimental techniques}
The experiment was performed at the Heavy Ion Research Facility of Lanzhou (HIRFL)~\cite{Zhan_NPA2008} in November 2017. A $^{32}$S$^{16+}$ primary beam was accelerated using the K69 Sector Focus Cyclotron and the K450 Separate Sector Cyclotron to 80.6~MeV/nucleon at an intensity of $\sim$87~$e$nA ($\sim$5.4~$p$nA). The secondary radioactive ions were produced via the projectile fragmentation of the $^{32}$S beam impinging on a 1581~$\mathrm{\mu}$m thick $^9$Be target. The main setting of the Radioactive Ion Beam Line in Lanzhou (RIBLL1)~\cite{Sun_NIMA2003} for the selection of the secondary beam was optimized on $^{27}$S. The average intensity and purity of $^{27}$S in the secondary beam delivered to the detection chamber were 0.14~particles per second (pps) and 0.024\%, respectively. The ions in the secondary beam were identified by the combination of energy loss ($\mathrm{\Delta}E$), time-of-flight (ToF), and magnetic rigidity ($B\rho$) according to the \textsc{lise}++ simulation~\cite{Tarasov_NIMB2008} and the calibration with the $^{32}$S primary beam. The ToF with respect to the two focus planes of RIBLL1 were given by two plastic scintillators (T1, T2), and the $\mathrm{\Delta}E$ was measured by two silicon detectors ($\mathrm{\Delta}E1$, $\mathrm{\Delta}E2$). The implanted events can be identified event by event over the entire experiment, allowing for a reliable quantification of the number of implanted $^{27}$S ions and thus facilitating the normalization of absolute proton and $\gamma$-ray intensities.

We designed a detection system composed of several double-sided silicon strip detectors (DSSD)~\cite{Xu_NST2018} and quadrant silicon detectors (QSD)~\cite{Bao_CPC2014}. Under a continuous-beam mode, the isotopes of interest were implanted into DSSD1 of 142~$\mathrm{\mu}$m thickness, DSSD2 of 40~$\mathrm{\mu}$m thickness, and DSSD3 of 304~$\mathrm{\mu}$m thickness in a certain proportion, where the subsequent decays were measured and correlated to the preceding implantations by using the position and time information. The three DSSDs were W1-type DSSD produced by the Micron Semiconductor Ltd.~\cite{Micron}. Thereinto, DSSD2 is the thinnest W1-type DSSD ever produced by the Micron Semiconductor Ltd, which was aimed at detecting low-energy protons because $\beta$ particles have a longer range in silicon, and accordingly, the $\beta$ particles emitted from a thin detector make small contributions to the background of the proton spectrum. DSSD3 has a higher detection efficiency for high-energy protons and $\beta$ particles, being an important supplement to the thinner DSSD2. Charged particles escaping from DSSD2 will deposit incomplete energies in DSSD2 and the residual energies of the escaping charged particles can be measured by DSSD1 or DSSD3 with high efficiency. The relationship between the energy-loss and the path of the escaping particles has proven to be a powerful method for light-particle identification~\cite{Xu_PLB2017}. A 1546~$\mathrm{\mu}$m thick QSD1 was installed downstream to detect the $\beta$ particles. QSD2 and QSD3, each with a thickness of $\sim$300~$\mathrm{\mu}$m, were installed at the end to veto the possible disturbances from the penetrating light particles ($^1$H, $^2$H, $^3$H, and $^4$He) coming along with the beam. In addition, five clover-type high-purity germanium (HPGe) detectors 
were employed to measure the $\gamma$ rays. All the silicon detectors were assembled compactly on printed circuit boards (PCBs), equipped with the SPA02- and SPA03-type preamplifiers developed on our own~\cite{Sun_NIMA2015}. This portable design has an advantage of easy customization for various experimental needs and facilitates the replacement and augmentation of detectors. The PCBs served as the mechanical support structure and also ensured that the detection system was properly grounded and shielded from sources of electromagnetic radiation. The PCBs inside the vacuum chamber were made of ceramic materials, which are good conductor of both electricity and heat, and will not decrease the degree of air vacuum by releasing their own molecules. The operating temperatures of silicon detectors and the preamplifiers were cooled to about $-2^{\circ}$C and $5^{\circ}$C, respectively, by using a cryogenic system. The low temperature was kept by a circulating cooling alcohol machine and be monitored by several digital thermistor thermometers. The cryogenic system dramatically reduced the leakage currents of the silicon detectors, which enabled us to suppress the intrinsic noise, to achieve a better resolution and to maintain the operation stability of the detection system. The DSSDs response for $\alpha$ particles was tested with a $^{241}$Am source, and a typical energy resolution of about 75~keV (FWHM) for the 5.486-MeV $\alpha$ particle was achieved for each strip on both sides of the DSSDs~\cite{Sun_NIMA2015}.

Each output channel of the preamplifiers for the three DSSDs was split into two parallel electronic chains with low and high gains in order to measure both the high-energy implantation events on the order of hundreds of MeV and the low-energy decay events with hundreds of keV or less. The logical OR signal of the three DSSDs was used to trigger the VME data acquisition (DAQ) system, which was a modified version of 'RIBF DAQ'~\cite{Baba_NIMA2010}. The dead time associated with present DAQ would reduce the accepted event rate. Taking the present experiment as an example, about 21.6\% of the decay events could not be recorded due to dead-time losses.

\section{Analysis and results}
\begin{figure}
\begin{center}
\includegraphics[width=6.6cm]{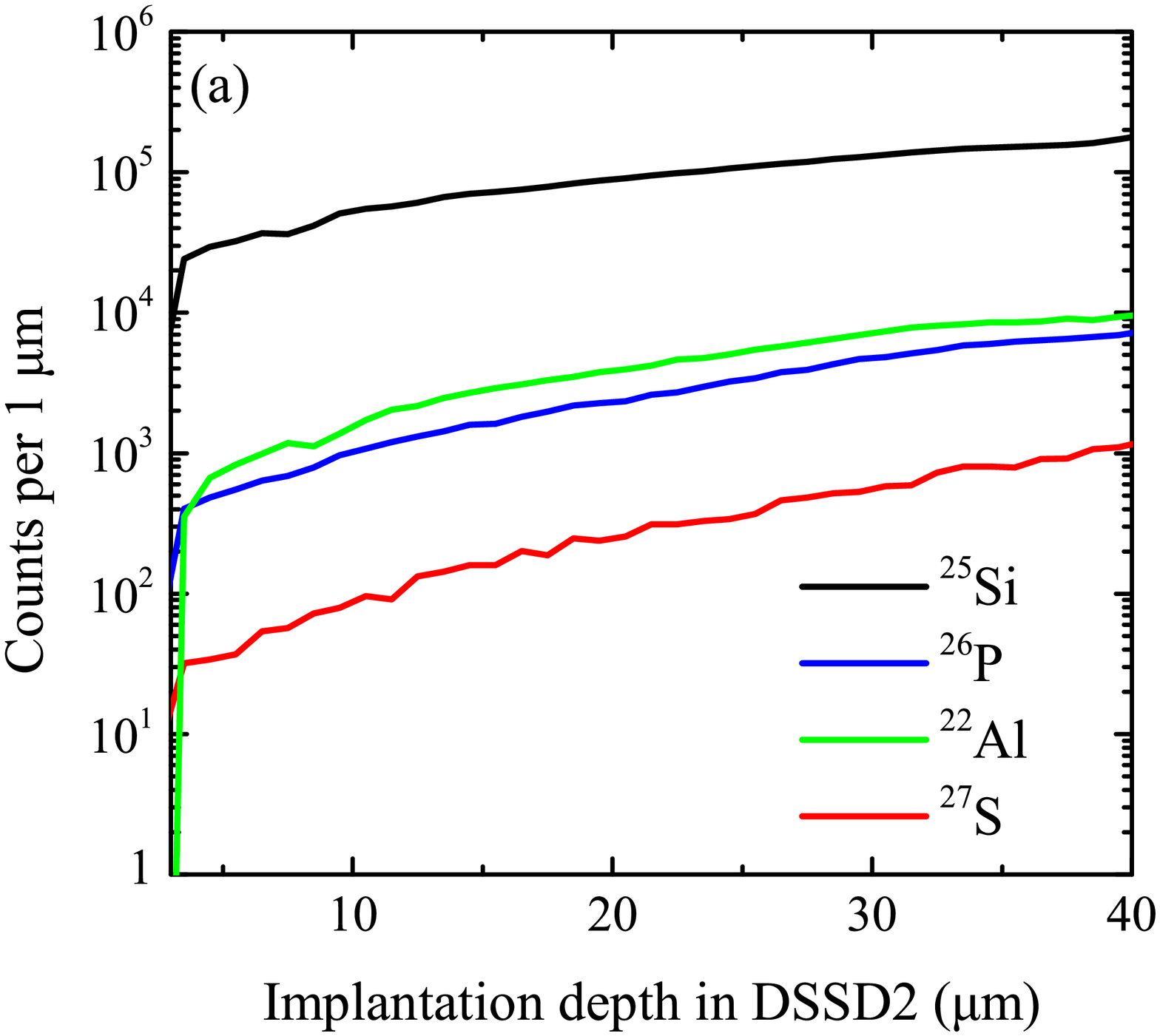}
\includegraphics[width=6.7cm]{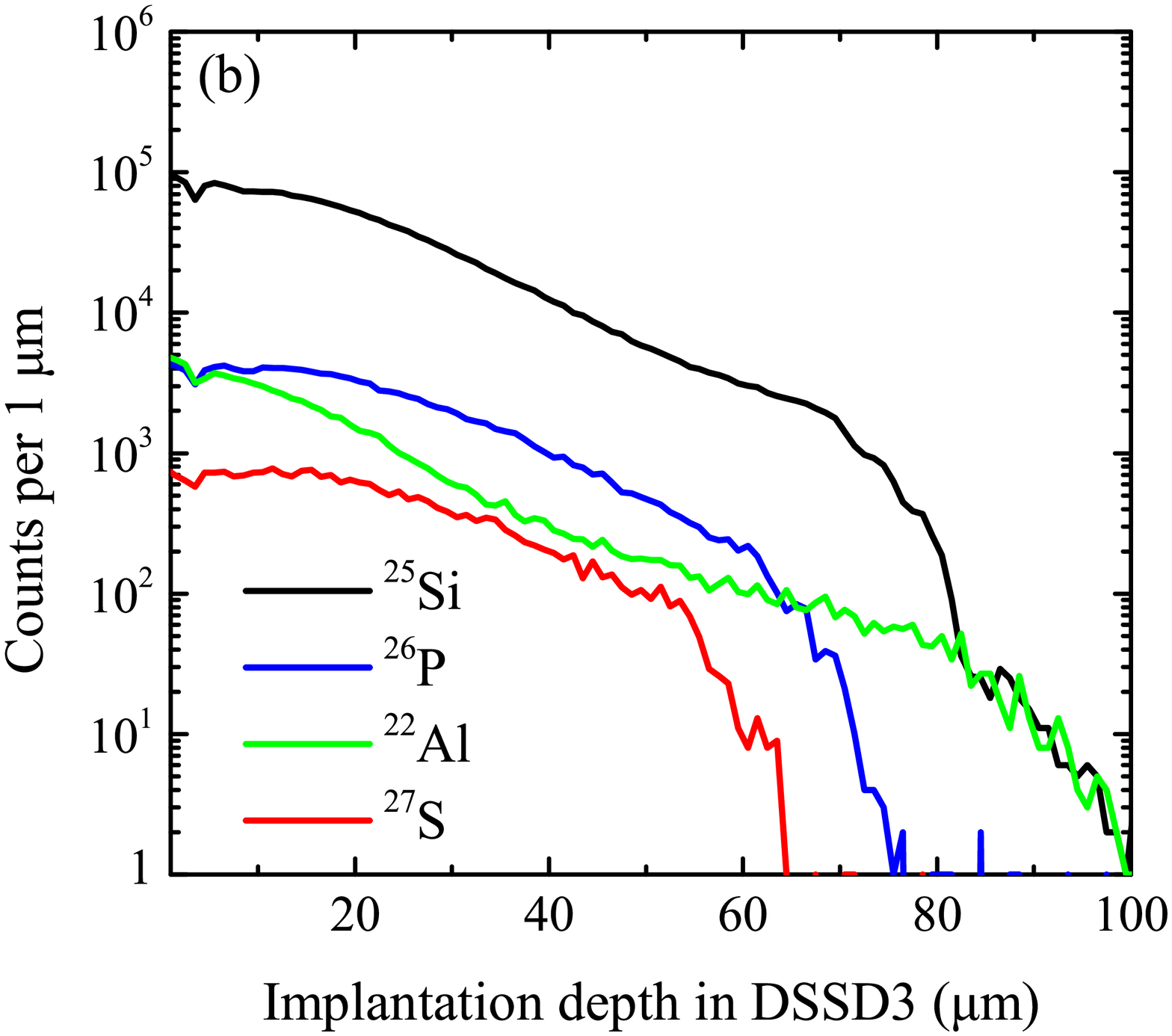}
\caption{\label{Implantation}Implantation depths distributions of $^{27}$S, $^{26}$P, $^{25}$Si, and $^{22}$Al ions measured by (a) DSSD2 and by (b) DSSD3.}
\end{center}
\end{figure}

The detection array was designed to maximize the efficiency of stopping the incoming heavy ions. A series of aluminum foils driven by three stepping motors were installed upstream to serve as a degrader. The thickness of the aluminum degrader could be adjusted with a small step and a full range of 416~$\mathrm{\mu}$m, so the stopping range of the ions in the DSSDs could be tuned accordingly. The beam was defocused to spread the ions on the surface of DSSD, resulting in a relatively low implantation rate in a single pixel. The stopping efficiency was estimated to be almost 100\% using silicon detectors with multiple thicknesses. The data were collected for 95.3~hours, excluding the time between each run. A total of $4.7\times10^4$~$^{27}$S ions were implanted into DSSD1, DSSD2, and DSSD3 with proportions of 0.6\%, 40.6\%, and 58.7\%, respectively. The present statistics on $^{27}$S is much higher than the statistics of $\sim1\times10^4$ achieved by Canchel \textit{et al}.~\cite{Canchel_EPJA2001} and that of 1267 achieved by Janiak \textit{et al}.~\cite{Janiak_PRC2017}, so the reliability of the determination or restriction of the nuclear structure information can be improved since the decay events can be observed with higher statistics in this work. In the secondary beam, the accompanying contaminants $^{26}$P and $^{25}$Si ions were provided with average intensities of 0.8~pps and 20.7~pps, and average purities of 0.16\% and 3.7\%, respectively. The decays of $^{23}$Si and its beam contaminants, $^{22}$Al and $^{21}$Mg, were measured with the same detection setup in the latter stage of the experiment. Qualitatively, the higher-statistics data on the well-known protons and $\gamma$ rays from those beam contaminants decays were also collected, which can be used as good calibration references to validate and to optimize the analysis program in obtaining accurate information from $\beta$-decay spectroscopy. Quantitatively, the calibration should also be corrected for the different implantation distributions in the DSSDs and the different $\beta$ decay $Q$ values of these nuclei. The surface implantation distribution of ions can be given by the DSSD pixels, and the implantation depth of each ion in a DSSD can be converted from their energy-loss measured by the detector itself using a \textsc{srim} code~\cite{Ziegler_NIMB2010}. The implantation depth distributions of the beam ions in the DSSDs are shown in Fig.~\ref{Implantation}.


\subsection{Proton energy and efficiency calibration}
The $\beta$-delayed proton peaks from $^{25}$Si decay with known energies and absolute intensities (in parentheses) of 401(1)~keV (4.75(32)\%), 943(2)~keV (1.63(20)\%), 1804(8)~keV (0.58(13)\%), 1917(2)~keV (2.24(21)\%), 2162(4)~keV (1.73(22)\%), 2307(4)~keV (1.57(21)\%), 3463(3)~keV (2.68(26)\%), 4252(2)~keV (9.54(66)\%), and 5624(3)~keV (2.39(20)\%)~\cite{Thomas_EPJA2004} were used for the energy and detection efficiency calibrations of the DSSDs. In order to assess the possible difference between the $^{27}$S of interest and the calibration references, a Monte-Carlo simulation using the \textsc{Geant}4 program~\cite{Agostinelli_NIMA2003} was performed by taking into account the features of the detector geometry, orientation, resolution, and threshold. Monoenergetic protons were emitted isotropically from the initial positions and interacted with the DSSD to produce an energy spectrum. The initial position can be obtained by randomly sampling the measured implantation distributions of the beam ions, since the
parent nucleus is stopped completely in the DSSD prior to its proton emission. Under the same constraints, the relative difference on the proton energy associated with varying implantation distributions and $\beta$ decay $Q$ values of nuclei was estimated to be $\leq1\%_{\rm 0}$~of a given proton energy. So an additional 1$\%_{\rm 0}$~uncertainty should be added in quadrature to other uncertainties on each peak energy to account for the nuclei variations. The relative difference for the proton peak integral with varying implantation distributions and $\beta$ decay $Q$ values of nuclei was found to be $\leq$1\%. Similarly, it is necessary to propagate this uncertainty through the peak-intensity analysis. These negligible differences associated with variations on nuclides can be understood by considering the remarkably similar implantation distributions shown in Fig.~\ref{Implantation}, as well as the fact that $\beta$ summing on proton energy has been demonstrated to be relatively insensitive to the $\beta$-decay $Q$ values~\cite{Meisel_NIMA2017}.

For both $^{25}$Si and $^{27}$S, the decay energy measured by the DSSD is a combination of the proton energy, the recoil energy of the heavy ion induced by the emitted proton, and the energy loss of the $\beta$ particle deposited in the detector. The proton peak would be shifted to higher energy due to the additional energy deposited in the DSSD from $\beta$ particles~\cite{Trinder_NPA1997,Czajkowski_NPA1998,McCleskey_NIMA2013}. The \textsc{Geant} simulation indicates that the proton peak shifts due to the $\beta$-summing effect for $^{25}$Si and $^{27}$S should be basically identical. The heavy ion loses a fraction of its energy to the silicon lattice instead of to ionization, leading to a pulse-height defect of the heavy recoil. Yet in the present case, the recoiling ions situated in the nearby region of the nuclide chart are of the similar atomic mass and the trivial difference between recoils results in a minor energy change compared to the above-mentioned $\beta$-summing effect~\cite{McCleskey_NIMA2013,Borrel_NPA1991}. To sum up, the proton peak energy shift due to the $\beta$-summing effect and the pulse-height defect of the recoiling particles can be considered to be automatically corrected in this calibration approach~\cite{Saastamoinen_PRC2011,Trinder_NPA1997,Gorres_PRC1992,Liu_PRC1998}. For the proton energy deposited in a DSSD, the signal read out from the junction-side strip ($E_x$) and that from the ohmic-side strip ($E_y$) should be approximately equal. In order to further suppress the $\beta$ background, $E_x$ and $E_y$ are limited within $\pm10\%$ as well as no more than $\pm296$ and $\pm112$~keV, respectively, for DSSD2 and DSSD3~\cite{Smirnov_NIMA2005,Wallace_PLB2012,McCleskey_PRC2016}. It is also noteworthy that the energy calibration from an internal source is more accurate than that from an external source, as it is necessary to make extra corrections involving the incident angle of particles and the thickness of the dead layer in the latter case~\cite{Buscher_NIMB2008}. The absolute detection efficiency for the protons emitted in the $\beta$ decay of $^{27}$S can be deduced from the efficiency curve fixed by the known $\beta$-delayed proton peaks of $^{25}$Si, assuming a uniform efficiency of the DSSDs for detecting the protons from $^{25}$Si and $^{27}$S decays. In order to verify our calibration and simulation, we derived the energy and the absolute intensity of the strongest $\beta$-delayed proton peak from $^{26}$P decay to be $E_p=$ 416(8)~keV and $I_p=$ 10.7(9)\%, respectively, from the weighted average of the results measured by DSSD2 and DSSD3 in this work. A comparison of the measured energy with the literature values of 412(2)~keV~\cite{Thomas_EPJA2004}, 426(30)~keV~\cite{Janiak_PRC2017}, and $414.9\pm0.6{\rm(stat)\pm0.3(syst)\pm0.6(lit)}$~keV~\cite{Bennett_PRL2013} shows a reasonably good agreement. The measured intensity agrees with a recent value of 10.4(9)$\sim$13.8(10)\%~\cite{Janiak_PRC2017}, while a higher intensity of 17.96(90)\% was reported based on the proton spectrum containing a large $\beta$ background~\cite{Thomas_EPJA2004}.

\subsection{$\gamma$-ray energy and efficiency calibration}
\begin{figure}
\begin{center}
\includegraphics[width=8cm]{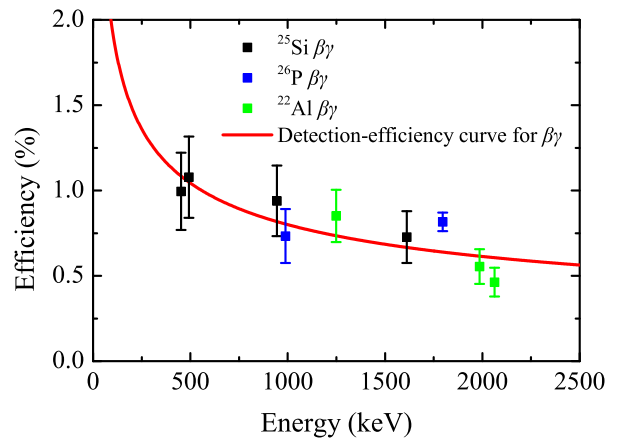}
\caption{\label{Efficiency}Absolute detection efficiency of the HPGe detectors for $\beta$-delayed $\gamma$ rays as a function of energy. This absolute efficiency is the product of the $\beta$-detection efficiency in DSSD3 and the $\gamma$-detection efficiency in the HPGe detectors. A fit function (red line): $\varepsilon=aE^b$ is used to parameterize the efficiency, where $\varepsilon$ is the detection efficiency at a given $\gamma$-ray energy $E$, and $a,b$ are free parameters.}
\end{center}
\end{figure}

Standard sources of $^{60}$Co, $^{137}$Cs, $^{133}$Ba, and $^{152}$Eu were placed at the exact centroid position of each DSSD after the detector array was removed from the vacuum chamber to calibrate the energy and intrinsic detection efficiency of the HPGe detectors. Four $\gamma$-ray transitions from the $\beta$-delayed $\gamma$ decay of $^{25}$Si with known energies and absolute intensities (in parentheses) of 452~keV (18.4(42)\%), 493~keV (15.3(34)\%), 945~keV (10.4(23)\%), and 1612~keV (15.2(32)\%)~\cite{Thomas_EPJA2004}, two $\gamma$-ray transitions from the $\beta$-delayed $\gamma$ decay of $^{26}$P with those of 988~keV (5.7(3)\%) and 1796~keV (58(3)\%)~\cite{Perez-Loureiro_PRC2016}, and three $\gamma$-ray transitions from the $\beta$-delayed $\gamma$ decay of $^{22}$Al with those of 1248.5(20)~keV (38.2(69)\%), 1985.6(13)~keV (31.1(54)\%), and 2062.3(15)~keV (34.1(58)\%)~\cite{Achouri_EPJA2006} were observed with high statistics in the present experiment. All nine of the $\gamma$-ray transitions were used together for the absolute detection efficiency calibration of the HPGe detectors. This absolute efficiency actually represents the efficiency to detect $\gamma$ rays in coincidence with $\beta$ particles measured by DSSD3. Though the $\beta$-delayed $\gamma$ decays of $^{21}$Mg and $^{23}$Si were also measured, they cannot be used as calibration references due to the lack of available results in literature~\cite{Lund_EPJA2015,Wang_IJMPE2018}. A \textsc{Geant}4 simulation including the implantation distributions and $Q$ values of relevant nuclei was also performed to estimate the efficiency of DSSD3 for detecting $\beta$ particles. Reductions of 1.1\%, 0.2\%, and 3.0\% in the $\beta$-detection efficiency for $^{25}$Si, $^{26}$P, and $^{22}$Al, respectively, were estimated compared with that for $^{27}$S. These values can be used as the normalization factors to correct the deviation between reference isotopes and $^{27}$S. As shown in Fig.~\ref{Efficiency}, the absolute efficiency for the $\gamma$ rays emitted in the $\beta$ decay of $^{27}$S can be deduced from the efficiency curve fixed by the nine known $\gamma$-ray lines, with the normalization factors for the $^{25}$Si, $^{26}$P, and $^{22}$Al decays taken into account. One of the sixteen crystals had worse resolutions and two of them were found to exhibit large gain drifts, so the data from these three crystals are discarded from the analysis.

To determine the centroids and the number of counts in each measured $\gamma$-ray peak or proton peak, the response function used to fit each peak is composed of a Gaussian function to describe the peak shape and a linear function to model the local background. In the case of peaks with multiple close contributions, a multipeak fitting function was applied to disentangle them. Statistical and systematic uncertainty are in general independent. In order to avoid an underestimation of the error, the statistical component and all the available systematic components of every uncertainty are supposed to be taken into account for the results determined in the following analysis sections.

\subsection{Half-life}
\begin{figure}
\begin{center}
\includegraphics[width=8cm]{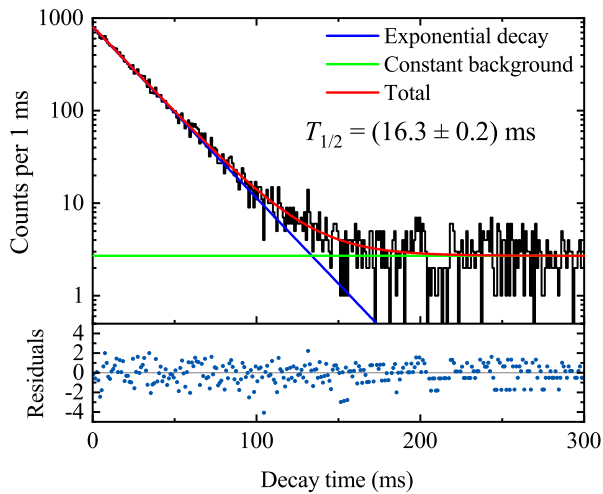}
\caption{\label{TimeSpec}Upper panel: decay-time spectrum of $^{27}$S. The spectrum is fitted with a formula (red line) which can be decoupled into an exponential decay component (blue line) and a constant background component (green line). Lower panel: the residuals of the fit divided by the square root of the number of events in each bin.}
\end{center}
\end{figure}

The half-lives of the nuclei along the rapid proton capture process pathway are important nuclear structure input for quantitative descriptions of explosive hydrogen burning in novae and XRBs. It is worthwhile to improve the precision of the lifetime of $^{27}$S, which represents a waiting point in XRBs nucleosynthesis~\cite{Parikh_PPNP2013}. As shown in Fig.~\ref{TimeSpec}, the decay-time spectrum of $^{27}$S was generated by the summation of the time differences between an implantation event and all the subsequent decay events which occur in the same $x$-$y$ pixel of the same DSSD. Multiple-pixels recorded decay events were rejected to reduce the probability for event mis-identifications, such as $\beta$ particles traveling along the detector through several pixels, or the rare events in which the ions were implanted very near the gap between strips, allowing the emitted protons to travel through two pixels. The decay-time spectrum contains a small quantity of random correlations, in which the implantation events could be accidentally correlated with decay events from other implantation events or disturbance events from background. All the true correlated implantation and decay event pairs generate an exponential curve whereas all the uncorrelated event pairs yield a constant background. A large time-correlation window between implantation-decay events could be achieved by the continuous-implantation method, which enabled us to accurately estimate the contribution of the background caused by uncorrelated events. In Fig.~\ref{TimeSpec}, a fit with a function composed of an exponential decay and a constant background yields the half-life of $^{27}$S to be $16.3\pm0.2$~ms. The uncertainty was directly derived from the fitting program, in which the half-life was treated as a free parameter and no preset parameters were involved. As can be seen from Table~\ref{T27S}, the present half-life of $^{27}$S is in good agreement with the theoretical prediction by the shell model. Our result is compatible with, and more precise than, all literature values~\cite{Borrel_NPA1991,Canchel_EPJA2001,Janiak_PRC2017}.

\begin{table}
\caption{\label{T27S}Half-lives of $^{27}$S.}
\begin{center}
\begin{ruledtabular}
\begin{tabular}{cc}
 Reference & $T_{1/2}$ (ms)\\
\hline
Borrel~\cite{Borrel_NPA1991}  & $21\pm4$\\
Canchel~\cite{Canchel_EPJA2001} & $15.5\pm1.5$\\
Janiak~\cite{Janiak_PRC2017}  & $15.5\pm1.6$\\
Present work  & $16.3\pm0.2$\\
Shell model  & 15.9 \\
\end{tabular}
\end{ruledtabular}
\end{center}
\end{table}

\subsection{$\beta$-delayed protons}
\begin{figure}
\begin{center}
\includegraphics[width=8.6cm]{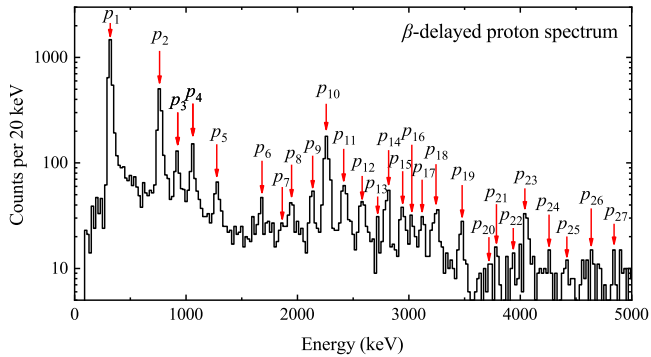}
\caption{\label{ProtonSpec}Cumulative $\beta$-delayed proton spectrum from $^{27}$S decay measured by DSSD2 and DSSD3. Each proton peak from the $\beta$-delayed proton decay of $^{27}$S is labeled with a letter $p$ followed by a number.}
\end{center}
\end{figure}

The cumulative $\beta$-delayed proton spectrum from $^{27}$S decay measured by DSSD2 and DSSD3 is shown in Fig.~\ref{ProtonSpec}, and each $\beta$-delayed proton peak from $^{27}$S decay is labeled with a letter $p$ followed by a number. The time differences between an implantation event and all the subsequent decay events were limited within about six half-life windows (96~ms). An anticoincidence with $\beta$-particle signals in QSD1 substantially suppresses the $\beta$-summing effect on the proton spectrum measured by DSSD3, which improves the overall energy resolution thus simplifying the identification of the proton branches. The corresponding intensity of each proton group can be calculated by the number of counts in the $\beta$-delayed proton peak in the spectrum, divided by the numbers of the implanted $^{27}$S ions given by the DSSDs. In this procedure, the subtraction of the proton peaks in the spectrum gated on the events within the constant background region in the decay-time spectrum (larger than ten half-life windows), the proton-detection efficiency correction of the DSSDs obtained from the $^{25}$Si efficiency calibration, and the dead-time correction of the DAQ system should be applied, as well. The dead-time correction was performed by using the trigger rate and the accepted counting rate, which had been recorded by the scalers in the DAQ system itself. The energy and the intensity for every proton group from $^{27}$S decay are obtained from the weighted average of the values from DSSD2 and DSSD3, except for $p_7$, $p_{20}$, and $p_{24-27}$ with too low intensities to be clearly identified on DSSD2. The uncertainty on energy was calculated through standard uncertainty propagation taking into account the statistical uncertainty on the peak centroid obtained from the fitting procedure (stat), the systematic uncertainty associated with the calibration parameters of the DSSDs (cali), the residuals of the calibration points with respect to the calibration line (resi), the systematic uncertainty associated with the adopted literature calibration references (lit), and the systematic uncertainty due to the nuclide variations obtained from the above-mentioned \textsc{Geant}4 simulation (simu). For example, the energies of the two strongest $\beta$-delayed proton peaks measured in the present work can be described as: $E_{p1}=318\pm0.3\rm(stat)\pm4.2(cali)\pm4.8(resi)\pm1.0(simu)\pm3.3(lit)$~keV and $E_{p2}=762\pm0.9\rm(stat)\pm4.3(cali)\pm4.8(resi)\pm1.0(simu)\pm3.3(lit)$~keV. Among the literature energies of the nine proton peaks used in the calibration~\cite{Thomas_EPJA2004}, the average uncertainty of 3.3~keV was adopted as the uncertainty of literature calibration references. Likewise, the uncertainty on intensity was calculated following the law of uncertainty propagation taking into account the statistical uncertainty on the peak areas obtained from the fitting procedure, the systematic uncertainties associated with the adopted literature calibration references, the residuals of the calibration points with respect to the calibration curve, and the systematic uncertainties associated with the nuclide variations obtained from the above-mentioned \textsc{Geant}4 simulation. In Table~\ref{bp27S}, the proton energies and intensities with available literature values are compared to the present results. As shown in Fig.~\ref{ProtonSpec}, the two strongest $\beta$-delayed proton peaks from $^{27}$S decay are marked with $p_1$ and $p_2$, which are identified as the only two proton peaks observed previously by Janiak \textit{et al}.~\cite{Janiak_PRC2017} based on their high intensities and energies, and a relatively better energy resolution has been achieved in the present work. The proton peak labeled with $p_{10}$ was previously observed by Canchel \textit{et al}.~\cite{Canchel_EPJA2001} with an energy of 2260(40)~keV and intensity of 1.9(4)\%. However, the energy and intensity of $p_{10}$ are measured to be 2264(9)~keV and 5.7(8)\%, respectively, in the present work. Janiak \textit{et al}.~\cite{Janiak_PRC2017} did not observe $p_{10}$ due to the type of detector, thus Canchel \textit{et al}.~\cite{Canchel_EPJA2001} is the only available measurement that can be used for comparison. A further measurement is required to resolve the discrepancies between the intensities reported in these two works. It is worth mentioning that the two-proton intensity obtained by Janiak \textit{et al}.~\cite{Janiak_PRC2017} was also larger by a factor of 3 than the result of Canchel \textit{et al}.~\cite{Canchel_EPJA2001}, which may support that our result of $p_{10}$ is more likely to be accurate. The half-lives of the three proton peaks are estimated to be 16.3(4), 16.6(8), and 16.7(11)~ms, respectively, which are consistent with the known half-life of $^{27}$S. The higher statistics of this experiment allowed us to estimate the half-life of each proton peak, providing further confirmation that the observed proton peaks did in fact originate from the $\beta$-delayed proton decay of $^{27}$S rather than from contaminants.

\begin{table*}
\caption{\label{bp27S}Decay energies ($E_p$) and intensities ($I_p$) for the $\beta$-delayed protons from $^{27}$S decay. A dash (--) indicates that no measurement was made of that quantity in that study.}
\begin{center}
\begin{ruledtabular}
\begin{tabular}{ccccccc}
Reference\footnotemark[1] & $E_{p1}$ (keV) & $I_{p1}$ (\%) & $E_{p2}$ (keV) & $I_{p2}$ (\%) & $E_{p10}$ (keV) & $I_{p10}$ (\%) \\
\hline
Togano~\cite{Togano_PRC2011}    & 315(17)   &  --  & 805(32)   & -- & -- & -- \\
Marganiec~\cite{Marganiec_PRC2016}    & 267(20)   & -- & 722(56)   & -- & -- & -- \\
Janiak~\cite{Janiak_PRC2017}    & 332(30)   & 24(3)$\sim$28(2)     & 737(30)   &  6.7(8)  & -- & -- \\
Canchel~\cite{Canchel_EPJA2001}    & --   & --   & --   &  --   & 2260(40) & 1.9(4) \\
Present work                              & 318(8)    & 23.1(21)   & 762(8)     & 8.9(10)  & 2264(9)   & 5.7(8)\\
\end{tabular}
\end{ruledtabular}
\footnotetext[1]{For the sake of completeness, the relative energies between $^{26}$Si and proton measured via the Coulomb dissociation of $^{27}$P from Refs.~\cite{Togano_PRC2011,Marganiec_PRC2016} are also listed. The relative energy is equivalent to the decay energy of the $\beta$-delayed protons from $^{27}$S decay.}
\end{center}
\end{table*}

\subsection{$\beta$-delayed $\gamma$ rays}
\begin{figure}
\begin{center}
\includegraphics[width=8cm]{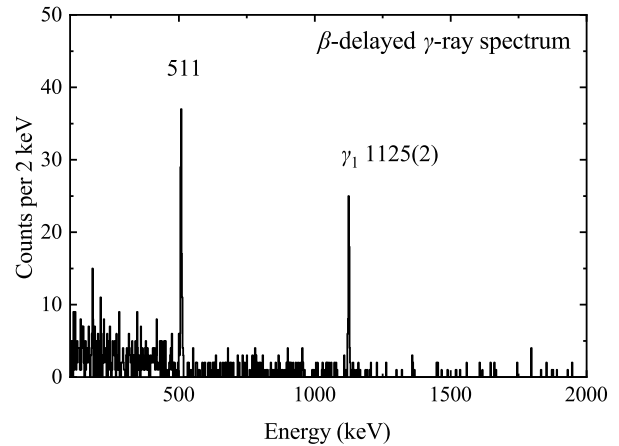}
\caption{\label{GammaSpec}Cumulative $\gamma$-ray spectrum measured by the HPGe detectors in coincidence with the $\beta$ particles from $^{27}$S decay measured by DSSD3. The $\gamma$-ray peak from the $\beta$-delayed $\gamma$ decay of $^{27}$S is labeled with its center-of-mass energy given in units of keV (with the error indicated in the parenthesis).}
\end{center}
\end{figure}

Figure~\ref{GammaSpec} shows the cumulative $\gamma$-ray spectrum measured by the HPGe detectors in coincidence with $^{27}$S $\beta$-decay signals in DSSD3. A $\beta$-delayed $\gamma$ ray at 1125(2)~keV is clearly observed for the first time in the $\beta$-decay measurements of $^{27}$S, which is the only statistically significant peak in the spectrum except for the well-known 511-keV $\gamma$ ray from the positron-electron annihilation. The uncertainty on the energy of the $\gamma$ ray was calculated following the law of uncertainty propagation taking into account the statistical uncertainty on the peak centroid obtained from the fitting procedure (stat) and the following systematic uncertainties. The uncertainty associated with the calibration parameters of the HPGe detectors (cali), the uncertainty associated with the adopted source data (src), and the residuals of the calibration points with respect to the calibration line (resi) were added in quadrature to obtain the total systematic uncertainty. The energy of the $\beta$-delayed $\gamma$ ray measured in the present work can be described as: $E_{\gamma1}=1125\pm0.5\rm(stat)\pm0.8(cali)\pm0.7(resi)\pm0.1(src)$~keV. Even the largest uncertainty on the $\gamma$-ray energy of the chosen sources have been determined to be on the order of tens of eV~\cite{Helmer_NIMA2000}, so a conservative estimation of the uncertainty (src) of 0.1~keV was adopted. Compared with the previously most precise energy of 1120(8)~keV from in-beam $\gamma$-ray spectroscopy~\cite{Gade_PRC2008}, an improvement in the uncertainty on this energy by a factor of 4 is obtained. The estimated half-life of this $\gamma$ ray is 16.1(24)~ms, consistent with that of $^{27}$S. The 1125-keV $\gamma$ ray is assigned as the deexcitation from the $3/2^+$ first excited state to the ground state of $^{27}$P. The intensity of 1125-keV $\gamma$ ray was tentatively predicted to be 36(3)\% by adopting the assumption that, except for the $3/2^+$ first excited state of $^{27}$P, all other excited states of $^{27}$P decay only via proton emission~\cite{Janiak_PRC2017}. In this work, the absolute $\gamma$-ray detection efficiency of the HPGe detectors for the $\gamma$ rays emitted in the $\beta$ decay of $^{27}$S at 1125~keV was estimated to be $\varepsilon_{1125}=0.76(19)\%$ by the detection-efficiency curve shown in Fig.~\ref{Efficiency}. The absolute intensity of the 1125(2)-keV $\gamma$ ray in the $\beta$ decay of $^{27}$S is determined to be $I_{\gamma1}=$ 31.1(86)\% using the number of counts in the 1125(2)-keV $\gamma$ ray peak and the counts of the implanted $^{27}$S ions in DSSD3. The background subtraction, the dead-time correction of the DAQ system, and the absolute detection efficiency correction of the HPGe detectors at 1125~keV were also applied, which is similar to that of proton intensities analyzed above. The statistical uncertainty on the numbers of decay events and implantation events, and the systematic uncertainty associated with the adopted literature calibration references and the residuals of the calibration points with respect to the calibration curve were combined in quadrature to determine the total uncertainty of the intensity.

\section{Discussion}
\subsection{Proton-$\gamma$-ray coincidence}
\begin{figure}
\begin{center}
\includegraphics[width=8cm]{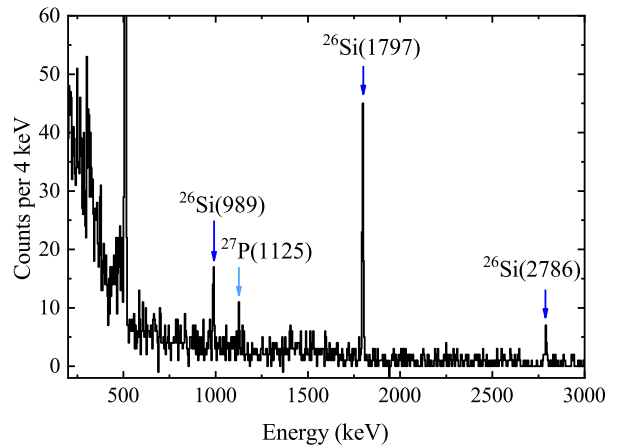}
\caption{\label{ProtongammaSpec}Cumulative $\gamma$-ray spectrum measured by the HPGe detectors in coincidence with the $\beta$-delayed protons from $^{27}$S decay measured by DSSD2 and DSSD3. The $\gamma$-ray peaks are labeled with their emitting nuclei and energies given in units of keV. The $\gamma$ rays at 989, 1797, and 2786~keV correspond to the deexcitations of the two lowest $2^+$ states of $^{26}$Si after the proton emissions from the excited states of $^{27}$P. The small contribution from the above-mentioned $\beta$-delayed $\gamma$ ray at 1125(2)~keV is also incorporated as the $\beta$ particles cannot be completely separated from the protons by DSSD.}
\end{center}
\end{figure}

In order to construct the decay scheme, it is necessary to apply a proton-$\gamma$-ray ($p$-$\gamma$) coincidence analysis. Figure~\ref{ProtongammaSpec} shows the $\gamma$-ray spectrum with a coincidence gating condition on protons from $^{27}$S decay. $\gamma$-ray peaks were observed at 989, 1797, and 2786~keV, which are attributed to the deexcitations from the two known excited states of $^{26}$Si following the proton emissions from the excited states of $^{27}$P. The half-lives of the 1797- and 2786-keV $2^+$ states were deduced to be 440(40) and 146(35)~fs, respectively~\cite{Basunia_NDS2016}. To get a qualitative understanding of the different origins of these three $\gamma$-ray peaks and the above-mentioned 1125-keV $\gamma$-ray peak, a simple Gaussian fit yields $\sigma_{989}=3.2$, $\sigma_{1797}=3.1$, $\sigma_{2786}=3.5$, and $\sigma_{1125}=2.3$. The former three peaks with larger $\sigma$ values, i.e. broad shapes, would likely have characteristic Doppler-broadened line shapes due to the $^{26}$Si recoil induced by the proton emission~\cite{Schwartz_PRC2015}. $p_1$ and $p_2$ were not observed in coincidence with any $\gamma$ rays, and hence they should be assigned as the proton emissions from the $3/2^+$ first excited state and the $5/2^+$ second excited state of $^{27}$P, respectively, to the ground state of $^{26}$Si. Due to the lack of $\gamma$-ray measurement, Canchel \textit{et al}. tentatively assigned $p_{10}$ as a proton emission to the ground state of $^{26}$Si~\cite{Canchel_EPJA2001}. However, in this work, $p_{10}$ is clearly observed in coincidence with the 989- and 1797-keV $\gamma$ rays and the efficiency-corrected number of $\gamma$ rays is compatible with the number of protons in $p_{10}$. Hence, it should be assigned as a proton emission to the second $2^+$ state at 2786~keV of $^{26}$Si. Likewise, the coincidence technique was systematically analyzed for all possible combinations of proton and $\gamma$ ray, and the results are summarized in Table~\ref{bp27Sall}. Though the $\gamma$ rays from the deexcitations of higher-lying $^{26}$Si states are not observed in the spectrum, it is possible that more $^{26}$Si states are populated by proton emissions. In this case, $p_1$ and $p_2$ contribute in large part to the total proton-emission intensity, and interpreting the higher-energy region of the proton spectrum is complicated due to their weak intensities. The protons and $\gamma$ rays were placed in the decay scheme based on spin and parity selection rules and their energy relationships, as well as including consideration of the decay scheme of mirror nucleus $^{27}$Na and our shell-model calculations.

\begin{table}
\caption{\label{bp27Sall}Decay energies ($E_p$) in the center-of-mass frame and absolute intensities ($I_p$) for the $\beta$-delayed protons from $^{27}$S decay, and the corresponding excitation energy of the initial state in $^{27}$P ($E^*_i$) and the final state in $^{26}$Si ($E^*_f$) for each transition measured in the present work. }
\begin{center}
\begin{ruledtabular}
\begin{tabular}{ccccc}
Proton & $E_{p}$ (keV) & $I_{p}$ (\%) & $E^*_i$ (keV) & $E^*_f$ (keV) \\
 \hline
$p_{1}$  & 318(8)   & 23.1(21) & 1125(2)  & 0 \\
$p_{2}$  & 762(8)   & 8.9(10)  & 1569(12) & 0 \\
$p_{3}$  & 913(9)   & 1.5(3)   & 4506(13) & 2786 \\
$p_{4}$  & 1054(9)  & 1.8(3)   & 1861(13) & 0 \\
$p_{5}$  & 1282(9)  & 1.1(2)   & 4875(13) & 2786 \\
$p_{6}$  & 1676(9)  & 0.6(2)   & 5269(13) & 2786 \\
$p_{7}$  & 1860(12) & 0.3(2)   & 4464(16) & 1797 \\
$p_{8}$  & 1951(11) & 0.8(2)   & 5544(15) & 2786 \\
$p_{9}$  & 2128(10) & 1.0(2)   & 5721(14) & 2786 \\
$p_{10}$ & 2264(9)  & 5.7(8)   & 5857(13) & 2786 \\
$p_{11}$ & 2417(11) & 1.6(4)   & 5021(15) & 1797 \\
$p_{12}$ & 2576(11) & 1.3(4)   & 6169(15) & 2786 \\
$p_{13}$ & 2717(10) & 0.6(2)   & 3524(14) & 0 \\
$p_{14}$ & 2808(10) & 2.0(5)   & 6401(14) & 2786 \\
$p_{15}$ & 2953(12) & 1.1(4)   & 6546(16) & 2786 \\
$p_{16}$ & 3030(12) & 1.0(3)   & 6623(16) & 2786 \\
$p_{17}$ & 3121(11) & 1.1(4)   & 6714(15) & 2786 \\
$p_{18}$ & 3238(11) & 1.4(4)   & 5842(15) & 1797 \\
$p_{19}$ & 3475(12) & 0.8(3)   & 7068(16) & 2786 \\
$p_{20}$ & 3720(11) & 0.4(2)   & 6324(15) & 1797 \\
$p_{21}$ & 3786(11) & 0.4(2)   & 7379(15) & 2786 \\
$p_{22}$ & 3950(11) & 0.4(1)   & 6554(15) & 1797 \\
$p_{23}$ & 4050(11) & 1.2(3)   & 6654(15) & 1797 \\
$p_{24}$ & 4260(15) & 0.4(2)   & 6864(18) & 1797 \\
$p_{25}$ & 4399(15) & 0.5(2)   & 7992(18) & 2786 \\
$p_{26}$ & 4693(15) & 0.4(2)   & 8286(18) & 2786 \\
$p_{27}$ & 4840(12) & 0.5(2)   & 7444(16) & 1797 \\
\end{tabular}
\end{ruledtabular}
\end{center}
\end{table}

\subsection{Decay scheme}
The $\beta$ decay of the $5/2^+$ ground state of $^{27}$S preferentially populates $J^\pi=3/2^+$, $5/2^+$, and $7/2^+$ states in $^{27}$P according to the $\beta$-decay selection rules. The $1/2^+$ ground state of $^{27}$P would be fed by the second-forbidden $\beta$ decay of the ground state of $^{27}$S, and therefore it would not contribute to the observed $\beta$ feedings. Typically, the $\beta$-feeding intensity to a bound state should be determined by subtracting the intensity of the $\gamma$ rays feeding this level from the intensity of the $\gamma$ rays deexciting this level, whereas in this case, all the excited states of $^{27}$P are proton unbound. The proton width is expected to be much larger than the $\gamma$ width for every state above the first excited states of $^{27}$P~\cite{Jung_PRC2012}. Therefore, the low-lying excited states of $^{27}$P are populated almost entirely by $\beta$ feeding rather than by $\gamma$ deexcitation from higher levels. The $\beta$-decay branching ratio to the $3/2^+$ first excited state of $^{27}$P populated in $^{27}$S decay is determined to be $I_{\beta1}=$~54.2(88)\% by the sum of the above-mentioned intensities of 1125(2)-keV $\beta$-delayed $\gamma$ ray and 318(8)-keV $\beta$-delayed proton in $^{27}$S $\beta$ decay, which is in agreement with the previous rough estimation of 60(4)$\sim$64(4)\%~\cite{Janiak_PRC2017}. The excitation energy of the first excited state in $^{27}$P was directly obtained from the measured $\gamma$-ray energy including a trivial correction (25~eV) for the energy carried by the daughter nucleus recoiling from $\gamma$-ray emission. The result compares fairly well with, as well as more precise than, the literature values of 1180~keV~\cite{Herndl_PRC1995}, 1199(19)~keV~\cite{Caggiano_PRC2001}, 1120(8)~keV~\cite{Gade_PRC2008}, 1176(32)~keV~\cite{Togano_PRC2011}, and 1137(33)~keV~\cite{Marganiec_PRC2016}. It should be noted that the excitation energies of the first excited state in $^{27}$P reported by Refs.~\cite{Togano_PRC2011,Marganiec_PRC2016} were deduced via a proton-separation energy of $^{27}$P of 861(27)~keV from AME2003~\cite{Audi_NPA2003} and a proton-separation energy of $^{27}$P of 870(26)~keV from AME2012~\cite{Wang_CPC2012}, respectively.

The proton-separation energy of $^{27}$P was deduced to be 807(9)~keV using the relation: $S_{p}(^{27}$P$)=E_{\gamma1}(^{27}$P~$3/2^+)-E_{p1}(^{27}$P~$3/2^+)$, where $E_{\gamma1}(^{27}$P~$3/2^+)$ equals to the excitation energy of 1125(2)~keV for the first excited state of $^{27}$P and $E_{p1}(^{27}$P~$3/2^+)$ is the proton-decay energy of 318(8)~keV corresponding to the proton emission from the first excited state in $^{27}$P to the ground state of $^{26}$Si. The present $S_{p}(^{27}$P) value is more precise compared with $S_{p}(^{27}$P$)=$ 870(26)~keV from AME2016~\cite{Wang_CPC2017} and $S_{p}(^{27}$P$)=$ 788(30)~keV reported by Janiak \textit{et al}~\cite{Janiak_PRC2017}.

Besides, the excitation energy of the $5/2^+$ second excited state in $^{27}$P was deduced to be 1569(12)~keV using the relation: $E^*(^{27}$P$~5/2^+)=S_{p}(^{27}$P$)+E_{p2}(^{27}$P~$5/2^+)$. This value is more precise than and consistent with those of 1592(62)~keV~\cite{Marganiec_PRC2016} and 1525(43) keV~\cite{Janiak_PRC2017}, while is slightly lower than those of 1660(40)~keV~\cite{Benenson_PRC1977}, 1615(21)~keV~\cite{Caggiano_PRC2001}, and 1666(42)~keV~\cite{Togano_PRC2011}. The $5/2^+$ second excited state was also predicted to have a five orders of magnitude larger proton-decay branch than $\gamma$-decay branch~\cite{Herndl_PRC1995}. No discernible $\gamma$-ray peak around the energy of 1569~keV can be observed in the $\beta$-delayed $\gamma$-ray spectrum presented in Fig.~\ref{GammaSpec}, and the corresponding $\beta$-decay branching ratio feeding this state is estimated to be $I_{\beta2}=$ 8.9(10)\% by using the measured intensity of the 762(8)-keV $\beta$-delayed proton emission from $^{27}$S decay. Likewise, the same analysis procedure was applied to determine the excitation energies and $\beta$-feeding intensities for all the other excited states of $^{27}$P, which are sufficiently high to decay primarily via proton emission, so each intensity of proton emission represents the $\beta$-decay branching ratio feeding the level. 

\subsection{Mass of $^{27}$P}
The proton emission and electromagnetic transition from the first excited state of $^{27}$P measured in the present work can be used as an alternative way to deduce the mass excess of the ground state of $^{27}$P. Combined with present $S_{p}(^{27}$P) value and the precise mass excesses of $^{26}$Si and $^{1}$H from AME2016~\cite{Wang_CPC2017}, the mass excess of the $^{27}$P ground state was deduced to be $-$659(9)~keV using the relation: $\mathrm{\Delta}(^{27}$P$)=\mathrm{\Delta}(^{26}$Si$)+\mathrm{\Delta}(^{1}$H$)-S_{p}(^{27}$P). The uncertainty on the mass excess of $^{27}$P is the quadrature sums of the uncertainties on the masses and energies adopted in the calculation. As shown in Table~\ref{Mass27P}, the present mass excess value is more precise than $\mathrm{\Delta}(^{27}$P$)=-$685(42)~keV recently measured via isochronous mass spectrometry~\cite{Fu_PRC2018}, $\mathrm{\Delta}(^{27}$P$)=-$640(30)~keV recently reported in Ref.~\cite{Janiak_PRC2017}, and $\mathrm{\Delta}(^{27}$P$)=-$722(26)~keV given by AME2016~\cite{Wang_CPC2017}. The AME2016 value was the weighted average of the two previously measured mass excesses of $-$753(35)~keV~\cite{Benenson_PRC1977} and $-$670(41)~keV~\cite{Caggiano_PRC2001}. The theoretical masses of $^{27}$P calculated by Bao \textit{et al}.~\cite{Bao_PRC2016}, Benenson \textit{et al}.~\cite{Benenson_PRC1977}, Schatz \textit{et al}.~\cite{Schatz_APJ2017}, and Fortune \textit{et al}.~\cite{Fortune_PRC2018} are also listed in Table~\ref{Mass27P} for comparison. It is expected that the mass uncertainty of $^{27}$P would affect the model predictions of XRB light curves strongly and also has a significant impact on the composition of the burst ashes~\cite{Schatz_APJ2017}, and the present result will provide a better constraint for modeling the nucleosynthesis in type I XRBs.

\begin{table}
\caption{\label{Mass27P}Comparison of the mass excesses of $^{27}$P obtained from the present work and from literature.}
\begin{center}
\begin{ruledtabular}
\begin{tabular}{ccc}
Reference & Method & $\mathrm{\Delta}(^{27}$P)~(keV) \\
\hline
Audi~\cite{Audi_NPA2003} & AME2003 & $-717\pm26$ \\
Wang~\cite{Wang_CPC2012} & AME2012 & $-722\pm26$ \\
Wang~\cite{Wang_CPC2017} & AME2016 & $-722\pm26$ \\
Bao~\cite{Bao_PRC2016} & Relation between mirror nuclei & $-779\pm290$ \\
Benenson~\cite{Benenson_PRC1977} & $^{32}$S$(^3$He$, ^8$Li$)^{27}$P & $-753\pm35$ \\
Benenson~\cite{Benenson_PRC1977} & Isobaric multiplet mass equation  & $-716\pm16$ \\
Caggiano~\cite{Caggiano_PRC2001} & $^{28}$Si$(^7$Li$, ^8$He$)^{27}$P & $-670\pm41$ \\
Schatz~\cite{Schatz_APJ2017} & Isobaric multiplet mass equation  & $-716\pm7$ \\
Janiak~\cite{Janiak_PRC2017} & $\beta$ decay of $^{27}$S & $-640\pm30$ \\
Fortune~\cite{Fortune_PRC2018} & Mirror energy differences & $-731$ \\
Fu~\cite{Fu_PRC2018} & Isochronous mass spectrometry & $-685\pm42$ \\
Present work & $\beta$ decay of $^{27}$S & $-659\pm9$ \\
\end{tabular}
\end{ruledtabular}
\end{center}
\end{table}

\subsection{Mirror asymmetry}
Comparison between the mirror decays also provides an opportunity to investigate the isospin asymmetry. $^{27}$S and $^{27}$Na is particularly interesting as an extension of this test. The degree of isospin-symmetry breaking can be quantified by the mirror-asymmetry parameter $\delta=ft^+/ft^--1$, where the $ft^+$ and $ft^-$ values are associated with the $\beta^{+}$ decay of $^{27}$S and the $\beta^{-}$ decay of $^{27}$Na, respectively. The $\beta$-decay energy of $^{27}$S was deduced to be 18337(78)~keV using the relation: $Q_{\mathrm{EC}}(^{27}$S$)=\mathrm{\Delta}(^{27}$S$)-\mathrm{\Delta}(^{27}$P), where the mass excess of $^{27}$P was determined above. The mass excess of $^{27}$S was estimated to be $\mathrm{\Delta}(^{27}$S$)=17678(77)$~keV using the Coulomb displacement energy systematics~\cite{Miernik_APPB2013} with the known mass excess of $\mathrm{\Delta}(^{25}$Al$)=-8915.97(6)$~keV~\cite{Wang_CPC2017} and the energy of the two-proton emission from the $^{27}$P isobaric analog state to the $^{25}$Al ground state, $E_{2p}=6372(15)$~keV, measured in this work. The mechanism of $\beta$-delayed two-proton emission of $^{27}$S requires a more complicated treatment, which is beyond the scope of the present paper and will be published in a forthcoming paper. With the $Q$ value, the half-life of $^{27}$S, the excitation energies and the $\beta$-feeding intensities to $^{27}$P levels measured in the present work, the corresponding log~$ft$ values for each $^{27}$P state can be calculated through the \textsc{logft} analysis program provided by the NNDC website~\cite{Gove_ADNDT1971}. The corresponding Gamow-Teller decay strengths, $B$(GT), were calculated from the $ft$ values using the following relation:
\begin{equation}
B(\text{GT})=\frac{K/g_V^2}{ft(g_A/g_V)^2}
\end{equation}
where $K/g_V^2=6144.2(16)$~s~\cite{Towner_RPP2010} and $(g_A/g_V)^2=(-1.2695(29))^2$~\cite{Yao_JPG2006}, with $g_V$ and $g_A$ being the free vector and axial-vector coupling constants of the weak interaction.

\begin{figure}
\begin{center}
\includegraphics[width=9cm]{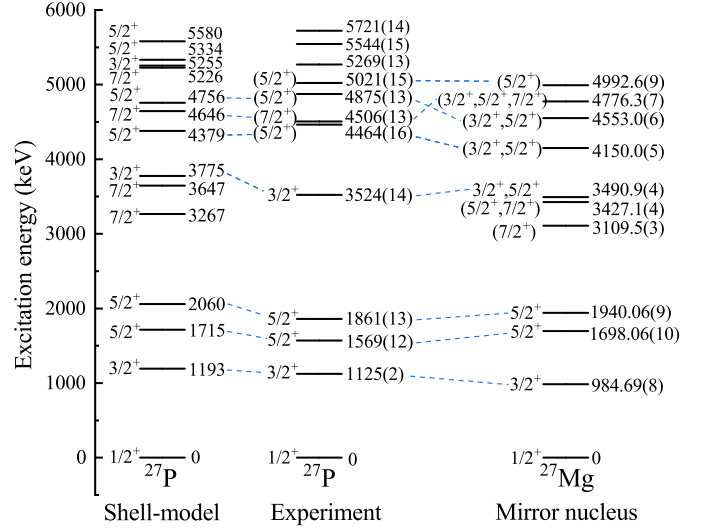}
\caption{\label{mirror_level} Low-lying $^{27}$P levels populated in $^{27}$S $\beta$ decay obtained in the present experiment and the $^{27}$Mg levels populated in $^{27}$Na $\beta$ decay from Ref.~\cite{Basunia_NDS2011}. The levels of $^{27}$P calculated by the shell model are shown for comparison.}
\end{center}
\end{figure}

Figure~\ref{mirror_level} shows the levels in $^{27}$P and $^{27}$Mg determined from the present measurement and the evaluation~\cite{Basunia_NDS2011}. This evaluation was based on two $\beta$-delayed $\gamma$-ray measurements of $^{27}$Na~\cite{Detraz_PRC1979,Guillemaud-Mueller_NPA1984}. The pairings for states above 4~MeV are matched by inference taking into consideration our theoretical calculations in the following section, and therefore should be taken with caution. The spectroscopic information of the mirror transitions and the mirror-asymmetry parameters extracted from the present measurement and the evaluation~\cite{Basunia_NDS2011} are reported in Table~\ref{Mirror27P}. The isospin asymmetry observed for the transition to the second excited state in the mirror $\beta$ decays of $^{27}$S and $^{27}$Na unambiguously confirms the assumption proposed by Janiak \textit{et al}.~\cite{Janiak_PRC2017}. A non-zero mirror-asymmetry parameter is sensitive to any abnormal nuclear structure in the initial and/or final state. Large mirror asymmetries have also been reported for transitions involving halo states~\cite{Perez-Loureiro_PRC2016,Borge_PLB1993,Ozawa_JPG1998}. It was shown that the weakly-bound effect of the proton $1s_{1/2}$ orbit contributed by Coulomb interaction might lead to halo states and also enhance the mirror asymmetries, but the potential relationship between large mirror asymmetries and halo structure of $sd$-shell proton-rich nuclei is still not entirely clear~\cite{Perez-Loureiro_PRC2016}. More systematic studies are needed in the future to better describe the contributions of possible effects that may produce mirror asymmetries.

\begin{table*}
\caption{\label{Mirror27P}Comparison between the excitation energies ($E^*$), $\beta$-feeding intensities ($I_\beta$), log~$ft$ values, and $B$(GT) values for the mirror transitions in the $\beta$ decays of $^{27}$S and $^{27}$Na. The mirror-asymmetry parameters $\delta$ are listed in the last column. The spin and parity assignments from Ref.~\cite{Basunia_NDS2011} are adopted. For the uncertain cases, the $J^\pi$ values favored by the present work are underlined.}
\begin{center}
\begin{ruledtabular}
\begin{tabular}{cccccccccc}
\multicolumn{4}{c}{$^{27}$S$\rightarrow^{27}$P $Q_{\mathrm{EC}}=18337(78)$~keV} &  & \multicolumn{4}{c}{$^{27}$Na$\rightarrow^{27}$Mg $Q_{\mathrm{\beta-}}=9010(40)$~keV~\cite{Basunia_NDS2011}}  & \\
\hline
$^{27}$P $E^*$ (keV)    & $I_\beta$ (\%)    & log~$ft$ & $B$(GT)   & $J^\pi$~\cite{Basunia_NDS2011}    & $^{27}$Mg $E^*$ (keV)    & $I_\beta$ (\%)    & log~$ft$ & $B$(GT) &  $\delta$ \\
\hline
1125(2) & 54.2(88) & 4.44(8) & 0.1384(256) & $3/2^+$ & 984.69(8) & 85.8  & 4.300 & 0.1911 & 0.38(26) \\
1569(12) & 8.9(10) & 5.16(5) & 0.0264(31) & $5/2^+$ & 1698.06(10) & 11.3  & 4.99  & 0.0390 & 0.48(18) \\
1861(13) & 1.8(3) & 5.82(8) & 0.0058(11) & $5/2^+$ & 1940.06(9) & 0.5   & 6.3   & 0.0019 & $-$0.67(7) \\
      &       &       &       & (\underline{$7/2^+$}) & 3109.5(3) & 0.5   & 5.91  & 0.0047 &  \\
      &       &       &       & ($5/2^+$,\underline{$7/2^+$}) & 3427.1(4) & 0.74  & 5.63  & 0.0089 &  \\
3524(14) & 0.6(2) & 6.04(14) & 0.0035(12) & \underline{$3/2^+$},$5/2^+$ & 3490.9(4) & 0.52  & 5.76  & 0.0066 & 0.91(62) \\
4464(16) & 0.3(2) & 6.24(21) & 0.0022(11) & ($3/2^+$,\underline{$5/2^+$}) & 4150.0(5)  & 0.026 & 6.81  & 0.0006 & $-$0.73(14) \\
4875(13) & 1.1(2) & 5.59(8) & 0.0098(19) & ($3/2^+$,\underline{$5/2^+$}) & 4553.0(6)  & 0.17  & 5.82  & 0.0058 & $-$0.41(11) \\
4506(13) & 1.5(3) & 5.50(7) & 0.0121(20) & ($3/2^+,5/2^+$,\underline{$7/2^+$}) & 4776.3(7) & 0.16  & 5.75  & 0.0068 & $-$0.44(10) \\
5021(15) & 1.6(4) & 5.39(9) & 0.0155(33) & (\underline{$5/2^+$}) & 4992.6(9) & 0.18  & 5.59  & 0.0098 & $-$0.37(14) \\
\end{tabular}
\end{ruledtabular}
\end{center}
\end{table*}

\subsection{Shell-model calculation}
We performed the theoretical calculations using the shell-model code \textsc{kshell}~\cite{Shimizu_arXiv2013} in the $sd$-shell model space involving the $\pi0d_{5/2}$, $\pi1s_{1/2}$, $\pi0d_{3/2}$, $\nu0d_{5/2}$, $\nu1s_{1/2}$, and $\nu0d_{3/2}$ valence orbits. The modified effective Hamiltonian (USD$^*$), including the shift of the single-particle energies and the reduction of the residual interaction related to the weakly-bound proton $1s_{1/2}$ orbit, was considered when applying the well-established shell-model Hamiltonian, USD~\cite{Brown_ARNPS1988}, to describe proton-rich weakly-bound nuclei. The reduction factors of the two-body matrix elements were evaluated by calculating the monopole-based universal interaction ($V_{\mathrm{MU}}$) in the Woods-Saxon basis~\cite{Yuan_PRC2014}. The configurations of the $1/2^+$ ground state, $3/2^+$ first excited state, and $5/2^+$ second excited state of $^{27}$P can be mainly described by the single-particle states of $\pi1s_{1/2}$, $\pi0d_{3/2}$, and $\pi0d_{5/2}$, respectively according to our calculation. The total experimental $\beta$-decay branching ratio to these twenty-seven states is determined to be $I_{\beta\rm{tot}}=$~91.0(90)\%. The value is compatible with the sum of theoretical branching ratios, $I_{\beta\rm{tot}}=$~91.4\%, excluding the four unobserved states. A quenching factor $q^2=0.55$ was used in the theoretical calculation. The observed $^{27}$P states have assignments of positive parities since they are most likely fed by allowed $\beta$ transitions. No negative parity states are obtained in the shell-model calculation as well as observed in the mirror nucleus $^{27}$Mg populated in the $\beta$ decay of $^{27}$Na.

As shown in Table~\ref{Shell27P}, every observed level up to 8286~keV is tabulated and matched with a specific theoretical level. We are not able to match the theoretical 7496- and 7990-keV states to the observed states. However, a discrepancy in low-energy region is that two $^{27}$P states at 3267 and 3647~keV predicted by the shell-model calculation are absent in our data. The two unbound $7/2^+$ states at 3267 and 3647~keV would act as $\ell=4$ proton resonances with respect to the $0^+$ $^{26}$Si ground state, where $\ell$ is the relative orbital quantum number of the proton with respect to the nucleus. In this case, the angular momentum transfer of the emitted proton $\ell = 4$ and therefore it should be strongly suppressed by the centrifugal barrier. The proton emissions from these two states are not energetically possible to feed the excited states of $^{26}$Si above 2302 and 2620~keV, respectively. Presumably, the proton emissions from two $7/2^+$ states populate the 1797-keV $2^+$ first excited state of $^{26}$Si, implying that the protons with energies of 663 and 1043~keV should be observed in coincidence with the 1797-keV $\gamma$ rays. However, no proton peak can be observed at 663~keV and the data also did not exhibit evidence for the 1797-keV $\gamma$ ray in coincidence with any protons around 663~keV within 60~keV, which is a typical 4$\sigma$-width for a proton peak. For the latter case, 1043~keV corresponds to the energy of $p_{4}$ in Fig.~\ref{ProtonSpec}, and one 1797-keV $\gamma$ ray is found to be in coincidence with one 1065-keV proton. The 989- and 1797-keV $\gamma$ rays correspond to the transitions feeding and deexciting the 1797-keV $2^+$ first excited state of $^{26}$Si, respectively. Taking into consideration the number of counts in $p_{4}$ (228.4), along with the number of counts and the intrinsic detection efficiency of the HPGe detectors for the 989-keV (33.5 and 2.37\%) and 1797-keV (86.3 and 1.56\%) $\gamma$ rays, approximately 4.9 $p_{4}$-1797 coincident events are expected to be observed. It is therefore inappropriate to assign a 3647-keV state of $^{27}$P with based on only one $p\gamma$ coincident event, and both of the 3267- and 3647-keV states are omitted from Table~\ref{Shell27P}. Every other state listed corresponds to an $\ell=2$ or $\ell=0$ proton emission to feed a $^{26}$Si state. All the $^{26}$Si and $^{27}$P states involved are positive parity states, whereas the $\ell=1$ proton emission requires a parity change. It is therefore not necessary to consider this scenario.

The general characteristics of the decay scheme measured in the present work including the excitation energies, $\beta$-feeding intensities, log~$ft$ values, and $B$(GT) values for the states of $^{27}$P can be reproduced well within the framework of the nuclear shell model taking the weakly-bound nature of the proton $1s_{1/2}$ orbit into consideration. Taking the astrophysically significant first excited state as an example, the calculated excitation energy of 1193~keV agrees with the measured value of 1125(2)~keV, but the calculation without the weakly-bound modifications would yield an excitation energy as low as 895~keV, which clearly shows the necessity of taking into account the weakly-bound nature of the proton $1s_{1/2}$ orbit. 

\begin{table*}
\caption{\label{Shell27P}Comparison of the experimental and theoretical excitation energies ($E^*$), $\beta$-feeding intensities ($I_\beta$), log~$ft$ values, and $B$(GT) values for the $^{27}$P states obtained in the present work. All the $^{27}$P states with a theoretical intensity larger than 0.1\% are listed. The total experimental $\beta$-decay branching ratios to the twenty-seven observed states is determined to be $I_{\beta\rm{tot}}=$~91.0(90)\%. This value is listed in the last row to compare with the sum of theoretical branching ratios, $I_{\beta\rm{tot}}=$~91.4\%, without taking the four unobserved states ($I_{\beta\rm{unobs}}=$~2.8\%) into account. The states above 4~MeV are matched tentatively, and their spin and parity assignments are indicated with the $J^\pi$ values in parentheses.}
\begin{center}
\begin{ruledtabular}
\begin{tabular}{ccccccccc}
 \multicolumn{4}{c}{Experiment} & & \multicolumn{4}{c}{Theory} \\
$E^*$ (keV)    & $I_\beta$ (\%)    & log~$ft$ & $B$(GT)   & $J^\pi$    & $E^*$ (keV)    & $I_\beta$ (\%)    & log~$ft$ & $B$(GT) \\
 \hline
1125(2)  & 54.2(88) & 4.44(8)  & 0.1384(256) & $3/2^+$  & 1193     & 56.6     & 4.34     & 0.177 \\
1569(12) & 8.9(10)  & 5.16(5)  & 0.0264(31) & $5/2^+$  & 1715     & 7.6      & 5.14     & 0.028 \\
1861(13) & 1.8(3)   & 5.83(7)  & 0.0056(10) & $5/2^+$  & 2060     & 0.2      & 6.62     & 0.001 \\
         &          &          &          & $7/2^+$  & 3267     & 0.6      & 6.02     & 0.004 \\
         &          &          &          & $7/2^+$  & 3647     & 1.9      & 5.45     & 0.014 \\
3524(14) & 0.6(2)   & 6.04(14) & 0.0035(12) & $3/2^+$  & 3775     & 1.8      & 5.45     & 0.014 \\
4464(16) & 0.3(2)   & 6.24(21) & 0.0022(11) & $(5/2^+)$  & 4379     & 1.0      & 5.60     & 0.010 \\
4506(13) & 1.5(3)   & 5.50(7)  & 0.0121(20) & $(7/2^+)$  & 4646     & 1.2      & 5.50     & 0.012 \\
4875(13) & 1.1(2)   & 5.59(8)  & 0.0098(19) & $(5/2^+)$  & 4756     & 2.4      & 5.16     & 0.027 \\
5269(13) & 0.6(2)   & 5.80(11)  & 0.0060(16) & $(7/2^+)$  & 5226     & 2.0      & 5.17     & 0.026 \\
5544(15) & 0.8(2)   & 5.62(9)  & 0.0091(19) & $(3/2^+)$  & 5255     & 0.2      & 6.21     & 0.002 \\
5857(13) & 5.7(8)   & 4.69(7)  & 0.0778(126) & $(5/2^+)$  & 5334     & 3.8      & 4.87     & 0.053 \\
5021(15) & 1.6(4)   & 5.39(9)  & 0.0155(33) & $(5/2^+)$  & 5580     & 2.0      & 5.10     & 0.031 \\
5721(14) & 1.0(2)     & 5.47(9)  & 0.0129(27) & $(3/2^+)$  & 5679     & 0.5      & 5.72     & 0.007 \\
5842(15) & 1.4(4)   & 5.30(10)  & 0.0191(45) & $(7/2^+)$  & 5696     & 0.1      & 6.27     & 0.002 \\
6169(15) & 1.3(4)   & 5.28(11) & 0.0200(51) & $(7/2^+)$  & 5978     & 1.7      & 5.10     & 0.031 \\
6324(15) & 0.4(2)   & 5.73(18) & 0.0071(30) & $(3/2^+)$  & 5993     & 1.2      & 5.23     & 0.023 \\
6401(14) & 2.0(5)     & 5.04(9)  & 0.0348(73) & $(7/2^+)$  & 6205     & 1.1      & 5.23     & 0.023 \\
6546(16) & 1.1(4)   & 5.28(12) & 0.0200(56) & $(7/2^+)$  & 6566     & 0.9      & 5.24     & 0.022 \\
6554(15) & 0.4(1)   & 5.76(13) & 0.0066(20) & $(7/2^+)$  & 6715     & 0.5      & 5.51     & 0.012 \\
6623(16) & 1.0(3)     & 5.30(10)  & 0.0191(45) &$(5/2^+)$  & 6837     & 1.0      & 5.16     & 0.027 \\
6654(15) & 1.2(3)   & 5.54(11) & 0.0110(28) & $(5/2^+)$  & 6916     & 0.6      & 5.33     & 0.018 \\
6714(15) & 1.1(4)   & 5.24(12) & 0.0219(61) & $(5/2^+)$  & 7089     & 1.9      & 4.81     & 0.059 \\
6864(18) & 0.4(2)   & 5.63(18) & 0.0089(38) & $(7/2^+)$  & 7273     & 0.5      & 5.39     & 0.016 \\
7068(16) & 0.8(3)   & 5.29(12) & 0.0196(55) & $(3/2^+)$  & 7290     & 1.4      & 4.90     & 0.049 \\
7379(15) & 0.4(2)   & 5.55(14) & 0.0107(35) & $(3/2^+)$  & 7476     & 0.2      & 5.63     & 0.009 \\
              &              &             &                  & $(7/2^+)$  & 7496     & 0.2      & 5.68     & 0.008 \\
7444(16) & 0.5(2)   & 5.45(16) & 0.0135(50) & $(5/2^+)$  & 7582     & 0.3      & 5.52     & 0.012 \\
7992(18) & 0.5(2)   & 5.31(18) & 0.0187(78) & $(3/2^+)$  & 7825     & 0.2      & 5.67     & 0.008 \\
              &             &               &                  & $(7/2^+)$  & 7990     & 0.1      & 5.89     & 0.005 \\
8286(18) & 0.4(2)   & 5.34(19) & 0.0174(77) & $(7/2^+)$  & 8405     & 0.5      & 5.08     & 0.032 \\
$I_{\beta\rm{tot}}$ & 91.0(90) &              &           &           &             & 91.4+2.8  &        &      \\
\end{tabular}
\end{ruledtabular}
\end{center}
\end{table*}

\subsection{Reaction-rate calculation}
The thermonuclear $^{26}$Si$(p,\gamma)^{27}$P reaction rate is the incoherent sum of all resonant and nonresonant capture contributions. As only resonances within the energy window contribute significantly to the reaction rate, the resonant part of the reaction rate ($N_A\langle\sigma\nu\rangle_r$) can be derived from the resonance energies and strengths of all resonances that are located in the effective energy windows, i.e. Gamow windows. For a narrow isolated resonance, the resonant reaction rate can be calculated using the well-known relation~\cite{He_PRC2017,Lam_APJ2016,Rolfs_1988},
\begin{equation}
\begin{split}
N_A\langle\sigma\nu\rangle_r=1.5394\times10^{11}(\mu T_9)^{-3/2}\times\omega\gamma\\
\times\mathrm{exp}\left(-\frac{11.605E_r}{T_9}\right)[\mathrm{cm^3s^{-1}mol^{-1}}]
\end{split}
\end{equation}
where $\mu=A_T/(1+A_T)$ is the reduced mass in atomic mass units, with $A_T=26$ as the mass number of $^{26}$Si. $E_r$ is the resonance energy in the center-of-mass system in units of MeV. $T_9$ is the temperature in units of Giga Kelvin (GK), and $\omega\gamma$ is the resonance strength in units of MeV, which is defined as:
\begin{equation}
\omega\gamma=\frac{2J_r+1}{2(2J_{T}+1)}\frac{\Gamma_p\times\Gamma_\gamma}{\Gamma_\mathrm{tot}}
\end{equation}
where $J_r$ is the spin of the resonance and $J_T=0$ is the spin of the ground state of $^{26}$Si. The total width $\Gamma_{\mathrm{tot}}$ of the resonance is the sum of its proton width ($\Gamma_p$) and $\gamma$ width ($\Gamma_\gamma$) since they represent the only two open decay channels for the resonances of relevance to the $^{26}$Si$(p,\gamma)^{27}$P reaction rate.

The Gamow energies and windows for $^{26}$Si$(p,\gamma)^{27}$P reaction shown in Table~\ref{Gamow} are calculated from a numerical study of the relevant energy ranges for astrophysical reaction rates~\cite{Rauscher_PRC2010}. At a temperature below 2~GK, the corresponding effective energy window is found to be below 1.11~MeV. There are only two observed resonances in $^{27}$P within 1.1~MeV above the proton-separation energy~\cite{Caggiano_PRC2001,Togano_PRC2011,Jung_PRC2012,Marganiec_PRC2016}, and their spins and parities have been determined unambiguously. For the key $3/2^+$ resonance at 318(8)~keV, its $\Gamma_p$ is calculated to be 2.55(74)~meV using the relation: $\Gamma_p=\Gamma_\gamma/(I_{\gamma}/I_p)$, where the experimental ratio of the $\gamma$-ray branch to the proton branch of $I_{\gamma}/I_p=1.35(39)$ is determined in the present work and the $\Gamma_\gamma=3.43(170)$~meV is adopted from Iliadis \textit{et al}.~\cite{Iliadis_NPA2010_3}. Combining the $\Gamma_\gamma$ and $\Gamma_p$ values yields an $\omega\gamma$ value of 2.92(191)~meV. Then the corresponding resonant reaction rate from the $3/2^+$ resonance contribution can be determined with our experimental data. The total width was calculated to 5.98(186)~meV using the relation: $\Gamma_{\rm{tot}}=\Gamma_p+\Gamma_\gamma$, so the lifetime was calculated to be 110(35)~fs using the relation: $\tau=\hbar/\Gamma_{\rm{tot}}$. The information about the key $3/2^+$ resonance needed to determine the reaction rate is summarized and compared to literature values~\cite{Herndl_PRC1995,Caggiano_PRC2001,Guo_PRC2006,Timofeyuk_PRC2008,Gade_PRC2008,Qi_SCPMA2009,Togano_NPA2005,Togano_EPJA2006,Iliadis_NPA2010_3,Togano_PRC2011,Jung_PRC2012,Fortune_PRC2015,Marganiec_PRC2016,Janiak_PRC2017} in Table~\ref{FirstRes27P}. The high and low values of the resonant reaction rate were derived from a 1 standard deviation variation of the resonance parameters. Alternatively, the $\Gamma_p$ was calculated to be $1.94^{+0.72}_{-0.52}$~meV with the Coulomb penetrability factor and the proton spectroscopic factor ($C^2S$) of the resonance, where the uncertainty from resonance energy was taken into account
~\cite{Iliadis_NPA1997}, providing an independent cross-check of the former $\omega\gamma$ value. The $C^2S$ corresponding to the proton capture to the $3/2^+$ final state was calculated to be 0.399 using the shell model with the above-mentioned USD interaction including the modification of the residual interaction related to the weakly-bound proton $1s_{1/2}$ orbit.

\begin{table}
\caption{\label{Gamow}Gamow windows $\widetilde E_{\rm hi}-\widetilde \Delta\leq E\leq\widetilde E_{\rm hi}$ and Gamow peaks $\widetilde E_0$ for the $^{26}$Si$(p,\gamma)^{27}$P reaction at a temperature $T$.}
\begin{center}
\begin{ruledtabular}
\begin{tabular}{cccc}
$T$~(GK) & $\widetilde E_{\rm hi}-\widetilde \Delta$~(MeV) & $\widetilde E_{\rm hi}$~(MeV) & $\widetilde E_0$~(MeV)  \\
  \hline
0.5 & 0.21 & 0.45 & 0.29  \\
1.0 & 0.25 & 0.67 & 0.39  \\
1.5 & 0.27 & 0.89 & 0.42  \\
2.0 & 0.28 & 1.11 & 0.48  \\
\end{tabular}
\end{ruledtabular}
\end{center}
\end{table}

\begin{table*}\scriptsize
\caption{\label{FirstRes27P}Resonance parameters of $3/2^+$ resonance in $^{26}$Si$(p,\gamma)^{27}$P reaction adopted in the present work and in literature.}
\begin{center}
\begin{ruledtabular}
\begin{tabular}{ccccccccc}
Reference & Method & $E^*$~(keV) & $E_r$~(keV) & $\Gamma_\gamma$~(meV) & $\Gamma_p$~(meV) & $\Gamma_\gamma/\Gamma_p$ & $C^2S$  & $\omega\gamma$~(meV) \\
  \hline
Herndl~\cite{Herndl_PRC1995} & shell model & 1180  & 320   & 1.36  & 1.7   & 0.8   & 0.414  & 1.51  \\
Caggiano~\cite{Caggiano_PRC2001} & $^{28}$Si$(^7$Li$, ^8$He$)^{27}$P & 1199(19) & 340(33) & 3.43  & 3.5   & 0.98  & 0.414 & 3.5  \\
Guo~\cite{Guo_PRC2006} & $^{26}$Mg$(d,p)^{27}$Mg ANC & 1199 & 338 & 3.43 & 12.7(12) & 0.27 &       & 5.4(1) \\
Timofeyuk~\cite{Timofeyuk_PRC2008} & microscopic cluster model &       &       &       & 4.04(77) &       & 0.48(5) &  \\
Gade~\cite{Gade_PRC2008} & one-proton knockout reaction & 1120(8) &       &       &       &       &       &  \\
Qi~\cite{Qi_SCPMA2009} & Skyrme-Hartree-Fock model & 1199 &       &   & 5.33  &  &       & 4.17  \\
Togano~\cite{Togano_NPA2005} & Coulomb dissociation of $^{27}$P &       &       & 0.36  &       &       &       &  \\
Togano~\cite{Togano_EPJA2006} & Coulomb dissociation of $^{27}$P &       &       & 1.3(8) &       &       &       &  \\
Iliadis~\cite{Iliadis_NPA2010_3} & compilation & 1120(8) & 259(28) & 3.4(17) & 0.180(72)  & 18.9(121)  & 0.60(15) & 0.34(28) \\
Togano~\cite{Togano_PRC2011} & Coulomb dissociation of $^{27}$P & 1176(32) & 315(17) & 0.096+1.2 & 4.04(77)  & 0.32  &  & $1.9^{+1.9}_{-1.1}$ \\
\multirow{2}[0]{*}{Fortune~\cite{Fortune_PRC2015}} & \multirow{2}[0]{*}{simple potential model} & \multirow{2}[0]{*}{} & \multirow{2}[0]{*}{315(17), 340(33)} & \multirow{2}[0]{*}{3.43, 1.2} & \multirow{2}[0]{*}{$2.3^{+2.1}_{-1.1}$, $6.0^{+11.4}_{-4.3}$} & \multirow{2}[0]{*}{} & \multirow{2}[0]{*}{0.60} & $1.58^{+0.31}_{-0.41}$, $2.00^{+0.25}_{-0.57}$,  \\
          &       &       &       &       &       &       &       & $2.77^{+1.10}_{-1.05}$, $4.36^{+1.37}_{-2.09}$  \\
Marganiec~\cite{Marganiec_PRC2016} & Coulomb dissociation of $^{27}$P & 1137(33) & 267(20) & 0.9499 & 0.5229 & 1.82  &       & 0.6745  \\
Janiak~\cite{Janiak_PRC2017} & $\beta$ decay of $^{27}$S & 1120(8) & 332(30) &       &       & 1.3(2)$\sim$1.5(2) &       &  \\
Present work & $\beta$ decay of $^{27}$S & 1125(2) & 318(8) & 3.43(170) & 2.55(74)  & 1.35(39) & 0.399 & 2.92(191) \\
\end{tabular}
\end{ruledtabular}
\end{center}
\end{table*}

\begin{table*}\scriptsize
\caption{\label{SecondRes27P}Resonance parameters of $5/2^+$ resonance in $^{26}$Si$(p,\gamma)^{27}$P reaction adopted in the present work and in literature.}
\begin{center}
\begin{ruledtabular}
\begin{tabular}{cccccccc}
Reference & Method & $E^*$~(keV) & $E_r$~(keV) & $\Gamma_\gamma$~(meV) & $\Gamma_p$~(eV) & $C^2S$  & $\omega\gamma$~(meV) \\
  \hline
Benenson~\cite{Benenson_PRC1977} & $^{32}$S$(^3$He$, ^8$Li$)^{27}$P & 1660(40)  &   &   &    &     &    \\
Herndl~\cite{Herndl_PRC1995} & shell model & 1660  & 800   & 0.33  & 13.61  &  0.126  & 0.99  \\
Caggiano~\cite{Caggiano_PRC2001} & $^{28}$Si$(^7$Li$, ^8$He$)^{27}$P & 1631(19) & 772(33) & 0.33  & 7.5  & 0.126 & 0.99  \\
Guo~\cite{Guo_PRC2006} & $^{26}$Mg$(d,p)^{27}$Mg ANC & 1631 & 770 & 0.33 & 17.4(16) &  & 0.99 \\
Qi~\cite{Qi_SCPMA2009} & Skyrme-Hartree-Fock model & 1631 &       &     & 17.7 &       & 0.99  \\
Iliadis~\cite{Iliadis_NPA2010_3} & compilation & 1631(19) & 772(33) & 0.33(17) & 4.3(17) &   & 0.66 \\
Togano~\cite{Togano_PRC2011} & Coulomb dissociation of $^{27}$P & 1666(42) & 805(32) &  &   &   & 0.60(11) \\
Marganiec~\cite{Marganiec_PRC2016} & Coulomb dissociation of $^{27}$P & 1592(62) & 722(56) & 0.1168 & 21.85  &       & 0.3504  \\
Janiak~\cite{Janiak_PRC2017} & $\beta$ decay of $^{27}$S & 1525(43) & 737(30) &       &       &       &  \\
Present work & $\beta$ decay of $^{27}$S & 1569(12) & 762(8) &  &   &   &  $0.65^{+0.34}_{-0.30}$ \\
\end{tabular}
\end{ruledtabular}
\end{center}
\end{table*}

\begin{table}
\caption{\label{Sfactor}Comparison of the $S$ factors at zero energy from literature.}
\begin{center}
\begin{ruledtabular}
\begin{tabular}{ccc}
Reference & Method & $S(0)$~(keV~b)  \\
  \hline
Herndl~\cite{Herndl_PRC1995} & shell model & 36.3  \\
Caggiano~\cite{Caggiano_PRC2001} & $^{28}$Si$(^7$Li$, ^8$He$)^{27}$P & 36.3  \\
Guo~\cite{Guo_PRC2006} & $^{26}$Mg$(d,p)^{27}$Mg ANC & 87(11) \\
Timofeyuk~\cite{Timofeyuk_PRC2008} & microscopic cluster model & 50(13)  \\
Iliadis~\cite{Iliadis_NPA2010_3} & compilation & 54.5 \\
Togano~\cite{Togano_PRC2011} & Coulomb dissociation of $^{27}$P & 21 \\
Marganiec~\cite{Marganiec_PRC2016} & Coulomb dissociation of $^{27}$P & 35.9  \\
Present & average & $47.5^{+50.5}_{-26.5}$  \\
\end{tabular}
\end{ruledtabular}
\end{center}
\end{table}

\begin{figure}
\begin{center}
\includegraphics[width=8.5cm]{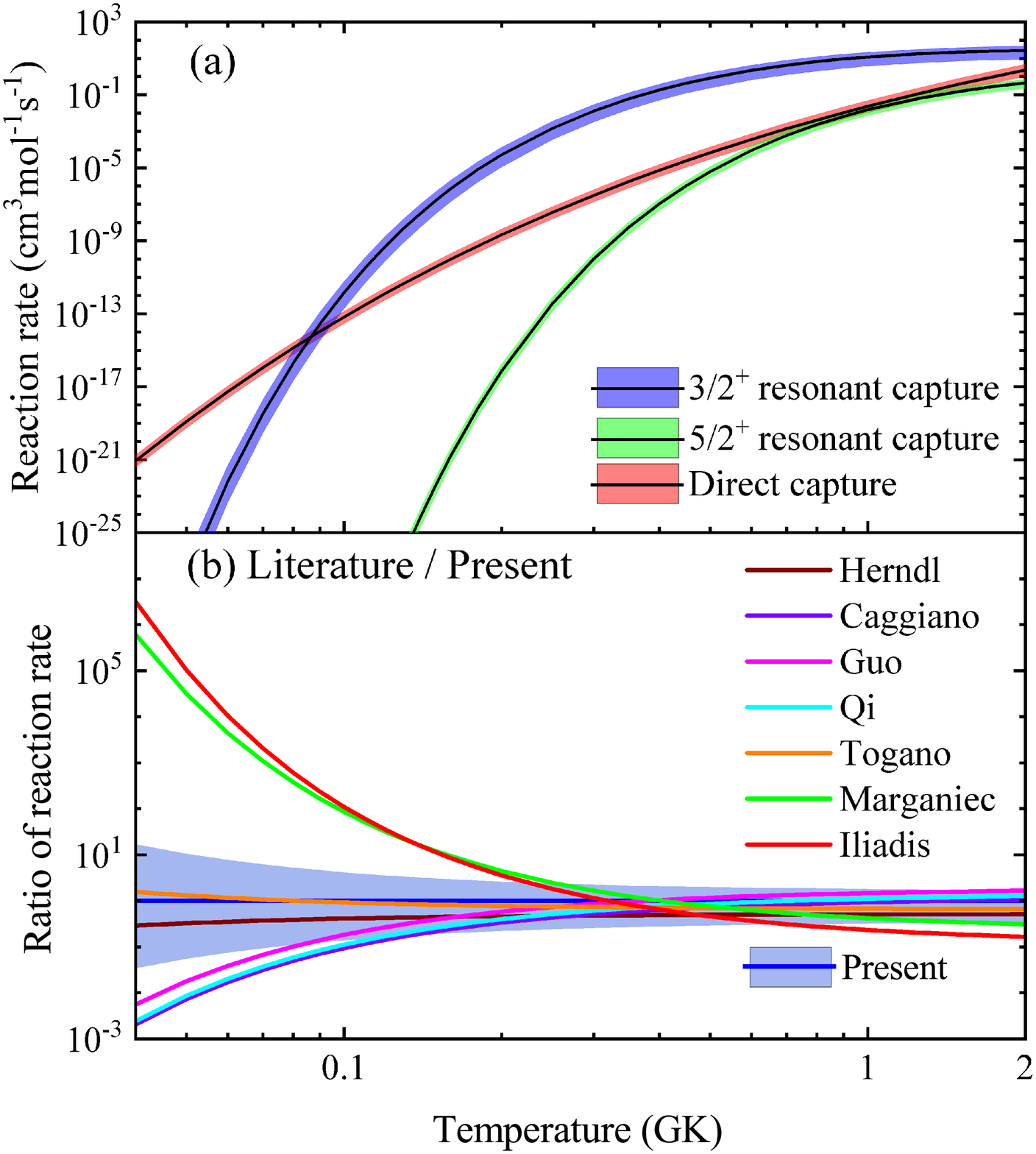}
\caption{\label{FirstReactionrate}Calculated thermonuclear $^{26}$Si$(p,\gamma)^{27}$P reaction rate. (a) Resonant-capture contribution from the $3/2^+$ resonance is determined with our experimental data. The average resonance strengths of $5/2^+$ resonance and the average zero-energy $S$ factors derived from all the available literature values are adopted in the calculations of the resonant-capture contribution from the $5/2^+$ resonance and direct-capture contribution, respectively. The high and low values of each reaction rate are shown in a colored band; (b) Ratios of the thermonuclear $^{26}$Si$(p,\gamma)^{27}$P reaction rate from the dominant $3/2^+$ resonance contribution from Herndl \textit{et al}.~\cite{Herndl_PRC1995}, Caggiano \textit{et al}.~\cite{Caggiano_PRC2001}, Guo \textit{et al}.~\cite{Guo_PRC2006}, Qi \textit{et al}.~\cite{Qi_SCPMA2009}, Iliadis \textit{et al}.~\cite{Iliadis_NPA2010_3}, Togano \textit{et al}.~\cite{Togano_PRC2011}, and Marganiec \textit{et al}.~\cite{Marganiec_PRC2016}, respectively, to that determined in the present work, including the present error band.}
\end{center}
\end{figure}

\begin{table*}\scriptsize
\caption{\label{Ratedata}Total thermonuclear $^{26}$Si$(p,\gamma)^{27}$P reaction rates in units of $\rm{cm^3s^{-1}mol^{-1}}$ obtained from the present work and from other literature.}
\begin{center}
\begin{ruledtabular}
\begin{tabular}{cccccccccc}
\multirow{2}[0]{*}{$T$ (GK)} & \multirow{2}[0]{*}{Herndl~\cite{Herndl_PRC1995}} & \multirow{2}[0]{*}{Caggiano~\cite{Caggiano_PRC2001}} & \multirow{2}[0]{*}{Guo~\cite{Guo_PRC2006}} & \multirow{2}[0]{*}{Iliadis~\cite{Iliadis_NPA2010_2}} & \multirow{2}[0]{*}{Togano~\cite{Togano_PRC2011}} & \multirow{2}[0]{*}{Marganiec~\cite{Marganiec_PRC2016}} & \multicolumn{3}{c}{Present work} \\
       &       &       &       &       &       &       & Low & Median & High \\
\hline
0.1   & $6.19\times10^{-13}$ & $2.67\times10^{-13}$ & $3.63\times10^{-13}$ & $1.38\times10^{-10}$ & $1.37\times10^{-12}$ & $1.21\times10^{-10}$ & $6.00\times10^{-13}$ & $1.32\times10^{-12}$ & $2.74\times10^{-12}$  \\
0.2   & $2.37\times10^{-05}$ & $2.98\times10^{-05}$ & $2.96\times10^{-05}$ & $1.46\times10^{-04}$ & $4.12\times10^{-05}$ & $2.30\times10^{-04}$ & $3.03\times10^{-05}$ & $4.57\times10^{-05}$ & $6.55\times10^{-05}$  \\
0.3   & $6.30\times10^{-03}$ & $1.16\times10^{-02}$ & $1.11\times10^{-02}$ & $1.24\times10^{-02}$ & $9.92\times10^{-03}$ & $2.19\times10^{-02}$ & $8.52\times10^{-03}$ & $1.16\times10^{-02}$ & $1.53\times10^{-02}$  \\
0.4   & $9.04\times10^{-02}$ & $2.02\times10^{-01}$ & $1.90\times10^{-01}$ & $1.06\times10^{-01}$ & $1.36\times10^{-01}$ & $1.88\times10^{-01}$ & $1.24\times10^{-01}$ & $1.63\times10^{-01}$ & $2.12\times10^{-01}$  \\
0.5   & $4.14\times10^{-01}$ & $1.04\times10^{+00}$ & $9.64\times10^{-01}$ & $3.51\times10^{-01}$ & $6.03\times10^{-01}$ & $6.33\times10^{-01}$ & $5.64\times10^{-01}$ & $7.38\times10^{-01}$ & $9.56\times10^{-01}$  \\
0.6   & $1.09\times10^{+00}$ & $2.95\times10^{+00}$ & $2.71\times10^{+00}$ & $7.38\times10^{-01}$ & $1.55\times10^{+00}$ & $1.35\times10^{+00}$ & $1.47\times10^{+00}$ & $1.92\times10^{+00}$ & $2.49\times10^{+00}$  \\
0.7   & $2.09\times10^{+00}$ & $5.99\times10^{+00}$ & $5.48\times10^{+00}$ & $1.21\times10^{+00}$ & $2.94\times10^{+00}$ & $2.25\times10^{+00}$ & $2.80\times10^{+00}$ & $3.66\times10^{+00}$ & $4.78\times10^{+00}$  \\
0.8   & $3.32\times10^{+00}$ & $9.92\times10^{+00}$ & $9.04\times10^{+00}$ & $1.70\times10^{+00}$ & $4.63\times10^{+00}$ & $3.20\times10^{+00}$ & $4.42\times10^{+00}$ & $5.80\times10^{+00}$ & $7.60\times10^{+00}$  \\
0.9   & $4.67\times10^{+00}$ & $1.44\times10^{+01}$ & $1.31\times10^{+01}$ & $2.17\times10^{+00}$ & $6.45\times10^{+00}$ & $4.13\times10^{+00}$ & $6.17\times10^{+00}$ & $8.12\times10^{+00}$ & $1.07\times10^{+01}$  \\
1.0   & $6.04\times10^{+00}$ & $1.90\times10^{+01}$ & $1.73\times10^{+01}$ & $2.62\times10^{+00}$ & $8.27\times10^{+00}$ & $4.99\times10^{+00}$ & $7.95\times10^{+00}$ & $1.05\times10^{+01}$ & $1.38\times10^{+01}$  \\
1.5   & $1.17\times10^{+01}$ & $3.90\times10^{+01}$ & $3.54\times10^{+01}$ & $4.41\times10^{+00}$ & $1.54\times10^{+01}$ & $8.02\times10^{+00}$ & $1.52\times10^{+01}$ & $2.00\times10^{+01}$ & $2.66\times10^{+01}$  \\
2.0   & $1.57\times10^{+01}$ & $5.06\times10^{+01}$ & $4.68\times10^{+01}$ & $6.65\times10^{+00}$ & $1.95\times10^{+01}$ & $1.05\times10^{+01}$ & $1.99\times10^{+01}$ & $2.58\times10^{+01}$ & $3.40\times10^{+01}$  \\
\end{tabular}
\end{ruledtabular}
\end{center}
\end{table*}

For the $5/2^+$ resonance in $^{27}$P, unlike the $3/2^+$ first excited state, the transition between the $5/2^+$ second excited state and the $1/2^+$ ground state of $^{27}$P is pure $E2$ multipolarity, so the $\omega\gamma$ of 0.60(11)~meV and $E_r$ of 805(32)~keV of the $5/2^+$ resonance were obtained directly from the relative energies and cross sections for the resonances measured via the Coulomb dissociation of $^{27}$P~\cite{Togano_PRC2011}. Recently, Marganiec \textit{et al}. reported a new $\omega\gamma$ of 0.3504~meV and $E_r$ of 722(56)~keV for the $5/2^+$ resonance with a similar approach~\cite{Marganiec_PRC2016}. Prior to these two measurements, a calculated $\omega\gamma$ of 0.99~meV~\cite{Herndl_PRC1995} had been widely adopted~\cite{Caggiano_PRC2001,Guo_PRC2006,Qi_SCPMA2009}. Such a tiny $\Gamma_\gamma$ value is expected to be far beyond the threshold of the sensitivity of the present experiment, nevertheless, the $E_r$ of the $5/2^+$ resonance was measured to be 762(8)~keV directly in our work, and an average $\omega\gamma$ value of $0.65^{+0.34}_{-0.30}$ was estimated by considering all the available previously determined $\omega\gamma$ values. By combining the average $\omega\gamma$ value and the present $E_r$ value of the $5/2^+$ resonance, the corresponding resonant reaction rate from the $5/2^+$ resonance contribution can be determined, however, its contribution to the total rate is negligible. The information about the $5/2^+$ resonance needed to calculate the reaction rate is summarized and compared to literature values in Table~\ref{SecondRes27P}. The high and low values of the resonant reaction rate were calculated from a 1 standard deviation variation of the resonance energy and from the lowest and highest known resonance strengths.

The nonresonant $^{26}$Si$(p,\gamma)^{27}$P reaction rate ($N_A\langle\sigma\nu\rangle_{\mathrm{dc}}$) is mainly determined by direct proton capture on the ground state of $^{26}$Si to the ground state of $^{27}$P since the ground state is the only proton-bound state of $^{27}$P. $N_A\langle\sigma\nu\rangle_{\mathrm{dc}}$ is directly related to the effective astrophysical $S$ factor ($S_{\rm eff}$) in the energy range of the Gamow window using the relation:
\begin{equation}
\begin{split}
N_A\langle\sigma\nu\rangle_{\rm{dc}}=7.8327\times10^{9}\left(\frac{Z_T}{\mu T_9^2}\right)^{1/3}\times S_{\rm eff}\\
\times\mathrm{exp}\left[-4.2487\left(\frac{Z_T^2\mu}{T_9}\right)^{1/3}\right][\rm{cm^3s^{-1}mol^{-1}}]
\end{split}
\end{equation}
where $\mu$ is the above-mentioned reduced mass in atomic mass units. $Z_T=14$ is the atomic number of $^{26}$Si. The $S$ factor is approximately constant over the range of the Gamow window outside the narrow resonances, so $S_{\rm eff}$ can be parameterized by the formula~\cite{He_PRC2017,Lam_APJ2016,Rolfs_1988,Fowler_ARAA1967},
\begin{equation}
S_{\rm{eff}}\approx S(0)\left[1+0.09807\left(\frac{T_9}{Z_T^2\mu}\right)^{1/3}\right]
\end{equation}
where $S(0)$ is the astrophysical $S$ factor at zero energy. All the available astrophysical $S$ factors from literature are listed in Table~\ref{Sfactor}, and the average of $S$ factors was estimated to be $S(0)=47.5$~keV~b, which was adopted in the calculation of the $^{26}$Si$(p,\gamma)^{27}$P reaction rate from direct-capture contribution. The direct capture rate only dominates the total rate below 0.08~GK. The high and low values of the nonresonant reaction rate were calculated from the largest and smallest known zero-energy $S$ factors, respectively.

Figure~\ref{FirstReactionrate}(a) displays the temperature dependence of the two resonant-capture contributions and the direct-capture component of the $^{26}$Si$(p,\gamma)^{27}$P reaction rate. Clearly, the $d$-wave ($\ell = 2$) radiative proton capture into the $3/2^+$ resonance in $^{27}$P makes the most dominant contribution to the total $^{26}$Si$(p,\gamma)^{27}$P reaction rate under a wide temperature range of 0.1$\sim$2~GK. The direct-capture component and the higher-lying resonances contributions are negligible when the temperature exceeds 0.08~GK. Figure~\ref{FirstReactionrate}(b) shows a comparison of the $^{26}$Si$(p,\gamma)^{27}$P reaction rate from the $3/2^+$ resonance contribution calculated by using the resonance energy and strength of the dominant $3/2^+$ resonance from the present work and those from literature. The rates differ from each other at temperatures below 0.3~GK. The deviation is enhanced mainly due to the differences in the adopted proton-separation energy of $^{27}$P and the resulting resonance energy of the $3/2^+$ resonance, especially for the evaluation~\cite{Iliadis_NPA2010_3}, in which an unreasonably low resonance energy of 259(28)~keV was adopted. As can be seen from Eq.~(2), the resonant reaction rate has an exponential dependence on the corresponding resonance energy and a linear dependence on the corresponding resonance strength.

Currently, the $^{26}$Si$(p,\gamma)^{27}$P reaction rate from Iliadis \textit{et al}.~\cite{Iliadis_NPA2010_2} recommended in both REACLIB~\cite{Cyburt_APJS2010} and STARLIB~\cite{Sallaska_APJS2013} databases, is universally adopted in nucleosynthesis studies. A comparison of the thermonuclear $^{26}$Si$(p,\gamma)^{27}$P reaction rates obtained from previous literature with that from the present work is shown in Table~\ref{Ratedata}. The low, median, and high present rates are calculated using the same Monte-Carlo technique from Iliadis \textit{et al}.~\cite{Iliadis_NPA2010_2} that incorporates our parameters of the key $3/2^+$ resonance. It is noteworthy that the present reaction rate is inconsistently smaller than that evaluated by Iliadis \textit{et al}.~\cite{Iliadis_NPA2010_2} at temperatures below 0.4~GK. The astrophysical impacts of our new $^{26}$Si$(p,\gamma)^{27}$P reaction rate were deduced via nova and XRB nucleosynthesis calculations~\cite{Jose_2016,Jose_APJ1998,Herwig_PoS2009,Koike_APJ2004}, which have been discussed in detail by Ref.~\cite{Sun_PRL2018}.

\section{Conclusion}
A detailed study of the $\beta$ decay of $^{27}$S was performed by using a continuous-implantation method. Simultaneous measurements of $\beta$-delayed proton and $\gamma$ decay of $^{27}$S were conducted for the first time. A total of twenty-seven $\beta$-delayed proton branches were identified, of which twenty-four have not been previously observed. A more complete $^{27}$S $\beta$-decay scheme was constructed with the experimental data including the accurate half-life of $^{27}$S, the excitation energies, $\beta$-feeding intensities, log~$ft$ values, and $B$(GT) values for the states of $^{27}$P obtained from the present work. We report the first experimental evidence for the observation of an 1125(2)-keV $\beta$-delayed $\gamma$ ray from $^{27}$S decay, corresponding to the exit channel of the astrophysically important $3/2^+$ resonance in $^{26}$Si$(p,\gamma)^{27}$P reaction. The excitation energy of 1125(2)~keV for the $3/2^+$ first excited state of $^{27}$P and the proton decay energy of 318(8)~keV from the first excited state in $^{27}$P to the ground state of $^{26}$Si also result in a most precise proton-separation energy (807(9)~keV) and the mass excess ($-$659(9)~keV) of $^{27}$P to date. The mirror-asymmetry parameters were deduced for eight transitions in the mirror $\beta$ decays of $^{27}$S and $^{27}$Na. The experimental spectroscopic information can be generally reproduced by the shell-model calculation taking the weakly-bound effect of the proton $1s_{1/2}$ orbit into account. More systematic investigations should be performed in the future to shed light on the possible effects that may give rise to mirror asymmetry~\cite{Kaneko_PLB2017}. With the nuclear physics input provided by this work, the $^{26}$Si$(p,\gamma)^{27}$P reaction rate is expected to be sufficiently well constrained for modeling the relevant nucleosynthesis in nova and type I XRB. Alternatively, the total width of the astrophysically important $3/2^+$ resonance can also be determined by the level lifetime. We provide a rough estimation of the lifetime of this resonance $\tau=110(35)$~fs, thus a lifetime measurement using the Doppler shift attenuation method in the future would be a beneficial complement to this work. Despite the fact that $p$-$\gamma$ coincidence analysis narrows down the various neutron-feeding possibilities and allows for a more confident determination of the decay scheme, not all the $p$-$\gamma$ coincidence can be clearly identified. Given the limited statistics of the present experiment, it is possible that the unobserved weak transitions lead to spurious assignments. It is particularly challenging to make definitive identifications regarding high-lying excited states of $^{27}$P where the density of states is quite high. We therefore propose our assignment as a tentative explanation of the high-lying levels feeding. The present work represents one step toward addressing these problems and parts of the $^{27}$S decay scheme still remain elusive. It would be great to perform independent experiments with higher statistics to approach a complete $^{27}$S decay spectroscopy with even forbidden $\beta$-decay transitions observed~\cite{Aboud_PRC2018,Fynbo_JPG2017}.

\begin{acknowledgments}
We acknowledge the continuous efforts of the beam physicists and operators of HIRFL for providing good-quality beams. We gratefully acknowledge Christian Iliadis for the reaction rate calculation. We would like to thank Zhihong Li and Christopher Wrede for the very helpful discussions. This work is supported by the National Key R\&D Program of China (2018YFA0404404 and 2016YFA0400503), and by the National Natural Science Foundation of China (11805120, 11805280, 11811530071, 11705244, 11705285, 11775316, U1732145, 11635015, 11675229, U1632136, 11505293, 11475263, 11490562, and U1432246), and by the China Postdoctoral Science Foundation (2017M621442 and 2017M621035), and by the International Postdoctoral Exchange Fellowship Program (Talent-Dispatch Program 20180068).
\end{acknowledgments}

\end{document}